\documentclass[useAMS,usenatbib,fleqn]{mn2e}

\usepackage[draft]{hyperref}
\usepackage{amssymb}
\usepackage{amsmath}
\usepackage{graphicx}

\title[Polarisation spectral synthesis for Type Ia supernova]{Polarisation spectral synthesis for Type Ia supernova explosion models}
\author[M. Bulla et al.]{M.~Bulla,$^1$\thanks{E-mail: mbulla01@qub.ac.uk} S.~A.~Sim,$^{1,2}$ M.~Kromer$^3$\\
$^1$Astrophysics Research Centre, School of Mathematics and Physics, Queen's University Belfast, Belfast BT7 1NN, UK\\
$^2$ARC Centre of Excellence for All-sky Astrophysics (CAASTRO)\\
$^3$The Oskar Klein Centre \& Department of Astronomy, Stockholm University, AlbaNova, SE-106 91 Stockholm, Sweden}
\date{Accepted, 23 March 2015. Received, 23 March 2015; in original form, 23 December 2014}

\usepackage{color}
\newcommand{\revised}[1]{#1}%{\textcolor{blue}{#1}}

\begin{document}

\maketitle 

\begin{abstract}
We present a Monte Carlo radiative transfer technique for calculating synthetic spectropolarimetry for multi-dimensional supernova explosion models. The approach utilises ``virtual-packets" that are generated during the propagation of the Monte Carlo quanta and used to compute synthetic observables for specific observer orientations. Compared to extracting synthetic observables by direct binning of emergent Monte Carlo quanta, this virtual-packet approach leads to a substantial reduction in the Monte Carlo noise. This is vital for calculating synthetic spectropolarimetry (since the degree of polarisation is typically very small) but also useful for calculations of light curves and spectra. We first validate our approach via application of an idealised test code to simple geometries. We then describe its implementation in the Monte Carlo radiative transfer code ARTIS and present test calculations for simple models for Type Ia supernovae. Specifically, we use the well-known one-dimensional W7 model to verify that our scheme can accurately recover zero polarisation from a spherical model, and to demonstrate the reduction in Monte Carlo noise compared to a simple packet-binning approach. To investigate the impact of aspherical ejecta on the polarisation spectra, we then use ARTIS to calculate synthetic observables for prolate and oblate ellipsoidal models with Type Ia supernova compositions.
\end{abstract}

\section{Introduction}

Type Ia supernovae (SNe Ia) are generally believed to be thermonuclear explosions of carbon-oxygen white dwarfs (see e.g. \citealt{roepke2011} and \citealt{hillebrandt2013} for reviews). However, answers to the questions of how and why the explosion is triggered remain unclear. Most of the established theoretical models involve close binary systems, but we still do not know if the companion star is a second white dwarf \citep[double degenerate system,][]{webbink1984,iben1984} or a non-degenerate star \citep[single degenerate system,][]{whelan1973}, and whether the explosion is triggered when an accreting white dwarf approaches the Chandrasekhar limit or via some other process. 

For Chandrasekhar-mass models, neither a pure deflagration nor a pure detonation model is able to fully account for the properties observed in SNe Ia: the former leads to strong turbulence and buoyancy resulting in fingers of nickel and carbon-oxygen at all ejecta velocities \citep{gamezo2003,roepke2006,jordan2012b,ma2013,fink2014} while the latter fails to produce intermediate-mass elements \citep{arnett1969}. However an interplay of these two models, so-called delayed-detonation models, remains promising for providing a good match to data. Possibilities include spontaneous deflagration to detonation transition models \citep{khokhlov1991,hoeflich1995,hoeflich1996a,hoeflich1996b,kasen2009,blondin2012}, and the gravitationally confined detonation model \citep{plewa2004,jordan2008,jordan2012a}. For non-Chandrasekhar-mass models, in which the accreting white dwarf may explode without approaching the Chandrasekhar limit, viable mechanisms are the detonation of helium layers on the surface of the accreting white dwarf \citep[the double detonation model; see][]{nomoto1980,woosley1980,livne1990,woosley1994,fink2007,fink2010,shen2009,woosley2011,moll2013} and the violent mergers of two white dwarfs \citep{pakmor2010,pakmor2012,pakmor2013,moll2014,raskin2014}. Alternative models include the explosion of white dwarf merger remnants \citep{benz1990,vankerkwijk2010,shen2012,kashyap2015}, the head-on collisions of white dwarfs  \citep{rosswog2009}, possibly induced in triple systems \citep{kushnir2013}, or the merger of a white dwarf with the hot core of an asymptotic giant branch star \citep{soker2014}.

There are a variety of ways to attempt to determine which of the proposed progenitor/explosion channels really occur (see e.g. \citealt{maoz2014} for a review). One approach is to perform explosion simulations with associated radiative transfer calculations and compare their predictions to data. Such work is well established \citep[e.g.][]{hoeflich1993,hauschildt1999,kasen2006,kromer2009,blondin2012,dessart2014,wollaeger2014} and it is now possible to compute synthetic observables from multi-dimensional models for a variety of explosion scenarios. Although state-of-the-art explosion simulations allow us to capture considerable complexity in SN Ia models (e.g. turbulence), degeneracies between models remain and make unambiguous interpretation difficult, even for the best observed nearby examples \citep{roepke2012}. 

One potentially powerful discriminant is the geometry, which can be quite different between models and depend on the nature of the progenitor and explosion mechanism. From the observational side, both nebular phase spectroscopy \citep{gerardy2007,maeda2010} and spectropolarimetric observations (see \citealt{wang2008} for a review) provide evidence that SNe Ia are not perfectly spherically symmetric. Continuum polarisation is typically quite low ($0.2-0.3$ per cent) in normal SNe Ia prior to optical maximum, pointing toward very small departures from global spherical symmetry; significant polarisation is, however, found across the line profiles of spectral features associated with intermediate-mass elements (calcium, silicon, sulphur and magnesium, but not oxygen; see e.g. \citealt{wang2008} and reference therein), suggesting that asymmetries in the element distribution are present. Other sub-classes of SNe Ia seem to display some peculiarities: sub-luminous SNe Ia show higher continuum polarisation levels \citep[$0.3-0.8$ per cent,][]{howell2001,patat2012}, whereas high-velocity SNe Ia show stronger line polarisation \citep[$\sim2$~per cent,][]{leonard2005,wang2006}. 

Asymmetric ejecta for SNe Ia are also predicted by many multi-dimensional explosion models, although with different degrees and types of asymmetry. For instance, the delayed-detonation model predicts that the ejecta can be quasi-spherical on large scales but with complex substructures \citep[in both density and composition,][]{seitenzahl2013,sim2013}, whereas simulations of violent white dwarf mergers predict departures from spherical symmetry on large angular scales \citep{pakmor2010,moll2014,raskin2014}. These differences in ejecta geometry between models have led to suggested connections with observed objects. For example, \citet{patat2012} proposed that the sub-luminous SN 2005ke might be explained by an explosion of a rotating white dwarf or a double-degenerate merger. It has also been suggested that several models could be ruled out because they yield explosions that are too aspherical and therefore inconsistent with continuum polarisation measurements for normal SNe Ia. For instance, \citet{maund2013} have claimed that the low continuum polarisation and significant line polarisation observed in SN 2012fr is consistent with delayed-detonation models but inconsistent with deflagration models or violent white dwarf mergers. 

However, interpreting polarisation data and quantifying arguments about its implications for models is difficult because estimating the degree of polarisation expected for a complex ejecta morphology (as provided e.g. by multi-dimensional explosion simulations) is not trivial: polarisation depends on the opacity distributions in a complex way \citep{hoeflich1991,dessart2011}. To date, spectropolarimetric data of SNe Ia have often been interpreted by comparing the observed polarisation levels with predictions from toy models with idealised configurations for the ejecta, e.g. ellipsoidal and clumped shell models or spherical shells with a hole or toroid \citep{howell2001,kasen2003,kasen2004,patat2012}. These idealised geometries are well-suited for building intuition and establishing the framework for interpreting polarisation data. However, quantitative comparisons between predictions of multi-dimensional explosion models and data require that polarisation calculations are made for the complete density/composition distributions predicted by hydrodynamic simulations. Such calculations make it possible to quantitatively compare the predictions of models to data (and each other) and assess the extent to which their geometries can really be distinguished via polarisation.

Here we present a polarisation scheme recently implemented in the three-dimensional, time-dependent Monte Carlo radiative transfer code ARTIS \citep[Applied Radiative Transfer In Supernovae,][]{sim2007,kromer2009}. The scheme involves two parts, first an implementation of Stokes parameters for the Monte Carlo quanta and secondly the development of techniques to reduce the Monte Carlo noise in the emergent synthetic observables. This is particularly important when we aim to extract very weak polarisation signals (low percentage levels are observed in SNe Ia), but also useful for total flux spectra and light curves. \revised{Although in this work we focus on the development, implementation and validation of the method using one-dimensional and two-dimensional models, our particular technique is well suited to exploit the multi-dimensional capability of ARTIS and therefore to be applied to three-dimensional explosion models}. Details of the methodology used are given in Section \ref{method} and the polarisation scheme is validated via an idealised test code in Section \ref{testcode}. We then present first results from the implementation in ARTIS, including testing with the one-dimensional W7 model \citep*{nomoto1984,iwamoto1999} and two-dimensional ellipsoidal toy models (Section \ref{artis}). We summarise and draw conclusions in Section \ref{conclusions}.

\section{Method}
\label{method}

\begin{figure}
\includegraphics[width=0.34\textwidth]{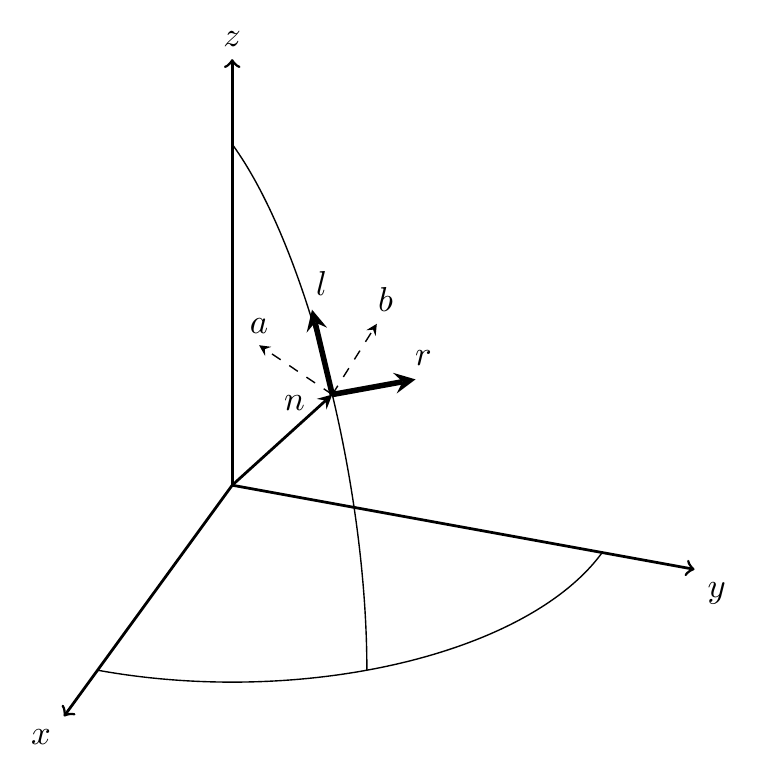}
\centering
\caption{The meridian plane coordinate system adopted in the Monte Carlo code. The Stokes vector is defined in the plane orthogonal to the direction of propagation, $\bmath{n}$. \revised{$Q$ is defined as the intensity difference between two perpendicular reference axes, $\bmath{l}$ and $\bmath{r}$, whereas $U$ is the equivalent difference with $\bmath{l}$ and $\bmath{r}$ counter-clockwise rotated by 45 degrees (viewed antiparallel to $\bmath{n}$).}}
\label{meridian}
\end{figure}

The polarisation of a beam of radiation is characterised by the four-dimensional Stokes vector $S = (I, Q, U, V)$. The first component, \textit{I}, gives the total intensity, \textit{Q} and \textit{U} measure the degree of linear polarisation and \textit{V} of circular polarisation. Since circular polarisation has never been observed in SNe Ia, and the radiative transfer calculations for circular and linear polarisation can be decoupled in scattering atmosphere in the absence of magnetic fields \citep{chandrasekhar1960}, here we neglect the $V$ component. The Stokes vector is defined in the plane orthogonal to the direction $\bmath{n}$ in which the radiation propagates. To define the Stokes parameters for linearly polarised radiation, we introduce two reference axes $\bmath{l}$ and $\bmath{r}$ so that $\bmath{l}$ lies in the meridian plane (plane defined by $\bmath{n}$ and the polar axis $z$) and $\bmath{n}=\bmath{r\times l}$ (see Fig. \ref{meridian}). With this convention, Q is defined as the difference between intensity $I_\text{l}$ with electric field oscillating along $\bmath{l}$ and intensity $I_\text{r}$ with electric field oscillating along $\bmath{r}$; U is the equivalent difference in intensities with the reference axes $\bmath{l}$ and $\bmath{r}$ counter-clockwise rotated by 45 degrees (as viewed looking antiparallel to $\bmath{n}$) to give $\bmath{a}$ and $\bmath{b}$. The resulting Stokes vector $S$ can be expressed as
\begin{equation}
S = \begin{pmatrix} I \\ Q \\ U \end{pmatrix} = \begin{pmatrix} I_\text{l}  + I_\text{r}  \\ I_\text{l} - I_\text{r} \\  I_\text{a} - I_\text{b} \end{pmatrix} = \begin{pmatrix}  \updownarrow + \leftrightarrow \\ \updownarrow - \leftrightarrow  \\ \mathrel{\rotatebox{45}{$\updownarrow$}}-\mathrel{\rotatebox{45}{$\leftrightarrow$}}  \end{pmatrix} ,
\end{equation}
or in terms of a dimensionless Stokes vector, $s$:
\begin{equation}
s = \frac{S}{I} = \begin{pmatrix}~ 1~ \\ q \\ u \end{pmatrix} .
\end{equation}
The polarisation fraction, $p$, and the position angle, $\chi$, of a beam are related to the Stokes parameters by
\begin{equation}
p = \frac{\sqrt{Q^2+U^2}}{I}=\sqrt{q^2+u^2} ~ ,
\end{equation}
\begin{equation}
\label{chi}
\chi = \frac{1}{2} \tan^{-1}{\bigg(\frac{U}{Q}\bigg)}= \frac{1}{2} \tan^{-1}{\bigg(\frac{u}{q}\bigg)} ~ ,
\end{equation}
where $\chi$ is the angle between the electric field orientation and the reference axis $\bmath{l}$. Spherically symmetric geometries are characterised by null polarisation since every contribution is canceled by an orthogonal contribution one quadrant away, whereas aspherical geometries may lead to a polarisation signal due to non perfect cancellation of the Stokes vectors (see also \citealt{kasen2003} and discussion in Section \ref{artis}).

\subsection{Propagation}
\label{propagation}
In this section, we adopt the terminology introduced \revised{by} \citet{lucy2002,lucy2005} and discuss the general scheme used to include polarisation in our radiative transfer code. The calculations we present use Monte Carlo methods: the radiative transfer problem is solved by simulating the propagation of Monte Carlo quanta (packets of identical photons) through an expanding medium. The propagation of an $r-$packet (monochromatic packet of ultraviolet-optical-infrared radiation) is followed through the ejecta in the rest frame (rf) and stopped by interactions with matter (which are treated in the comoving frame, cmf). For ultraviolet-optical-infrared radiation, the code currently accounts for both line opacity (treated in the Sobolev approximation [\citealt{sobolev1960}]) and continuum opacity due to electron scattering, bound-free and free-free absorption. To choose when continuum and interaction events occur we use the method outlined by \citet{mazzali1993}. \revised{Once a random optical depth $\tau_\text{r}$ is drawn, we determine the trajectory point at which the $r-$packet interacts with the next line (using the Sobolev approximation): if the continuum opacity accumulated up to that point is greater than $\tau_\text{r}$, a continuum absorption is selected; if instead the sum of continuum and line opacity is greater than $\tau_\text{r}$, a line event occurs; otherwise, the process is repeated for the next line with which the packet comes into resonance.} In an electron scattering event, the $r-$packet keeps the same cmf frequency and is assigned a new direction of propagation. For all other interactions (line absorption, free-free absorption and bound-free absorption) either a $k-$packet (packet of thermal kinetic energy) or \revised{an} $i-$packet (packet of excitation/ionisation energy) is activated and then processed according to the scheme proposed by \citet{lucy2002} as described by \citet{kromer2009}.

Our polarisation scheme adopts a method similar to that proposed by \revised{\citet{lucy2005}} and already implemented by \citet{kasen2006}. Polarisation is introduced by assigning a Stokes vector to each $r-$packet. When an interaction occurs, the Stokes parameters are transformed through the following sequence of steps: first, we transform the incoming Stokes vector $s_\text{i}$ in the rf to $s_\text{i}^\prime$ in the cmf~\footnote{In the following, cmf quantities are denoted with a prime, whereas unprimed quantities refer to the rf. }. The plane in which the electric field oscillates changes as a result of the aberration of the direction $\bmath{n}$ to 
\begin{equation}
\label{nprime}
\bmath{n^\prime} =  \frac{ \Bigg[ \bmath{n} - \frac{\bmath{v}}{c} \Bigg( \gamma-\frac{\gamma^2}{\gamma+1} \frac{\bmath{n \cdot v}}{c} \Bigg)  \Bigg]}{\gamma \Big( 1 - \frac{\bmath{n \cdot v}}{c} \Big) } ~ ,
\end{equation}
where $c$ is the speed of light, $\gamma$ the Lorentz factor and $\bmath{v}$ the local velocity of the ejecta \citep{castor1972}. To derive how the Stokes parameters change under the Lorentz transformation, we introduce a unit vector $\hat{\bmath{e}}$ that describes the orientation of the net electric field in the rf,
\begin{equation}
\hat{\bmath{e}} = ( \cos{\chi} ) ~\bmath{l} - (\sin{\chi})~ \bmath{r} ~ ,
\end{equation}
with the angle $\chi$ between $\hat{\bmath{e}}$ and $\bmath{l}$ computed from the incoming Stokes vector following equation (\ref{chi}). From this, it is easy to obtain a cmf Stokes vector representation that is relative to axes $\bmath{l^\prime}$ and $\bmath{r^\prime}$ (defined by $\bmath{n^\prime}$), making use of the fact that the polarisation, $p$, is invariant \citep{cocke1972}:
\begin{equation}
q_\text{i}^\prime = p_\text{i} \cos{2\chi^\prime}  \hspace{1cm} u_\text{i}^\prime = p_\text{i} \sin{2\chi^\prime} ~ .
\end{equation}
\vspace{0.01cm}

\begin{figure}

\includegraphics[width=0.34\textwidth]{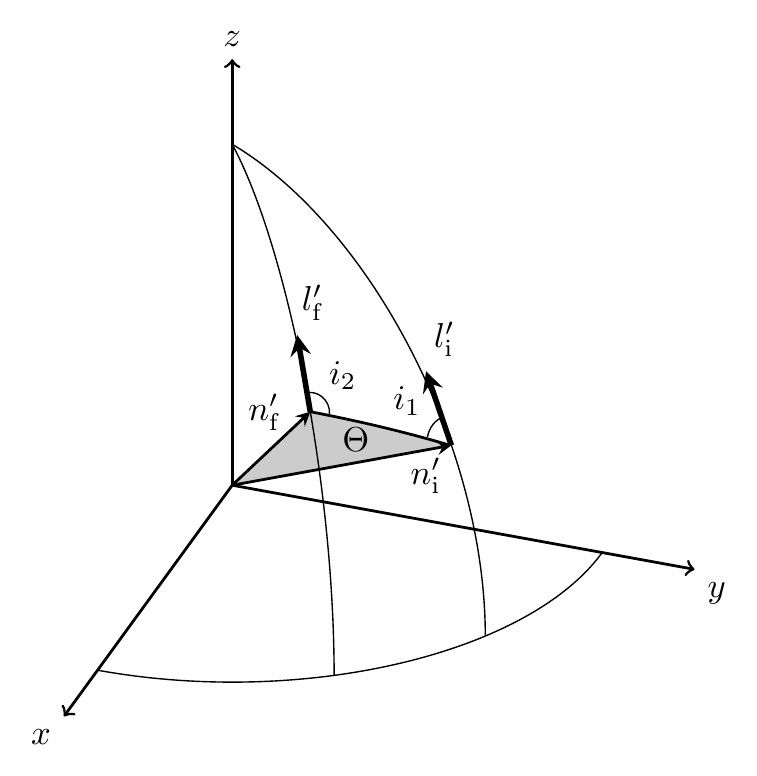}
\centering
\caption{The geometry adopted for electron scattering \revised{in the coordinate system introduced by \citet{chandrasekhar1960} with the additional reference axes $\bmath{l_\text{i}^\prime}$ and $\bmath{l_\text{f}^\prime}$ defined in the text}. A packet moving along $\bmath{n_\text{i}^\prime}$ is scattered through an angle $\Theta$ to a new direction $\bmath{n_\text{f}^\prime}$. The reference axis $\bmath{l_\text{i}^\prime}$ (and accordingly the Stokes vector) are rotated through an angle $i_1$ into the scattering plane (in grey) and through an angle $i_2$ after scattering.}
\label{scattering}
\end{figure}

Following transformation to the cmf, the Stokes parameters can be updated in accordance with the physical interaction that occurs. For \textit{bound-bound}, \textit{bound-free} and \textit{free-free} absorptions, we first activate either \revised{an} $i-$packet or a $k-$packet, as outlined in \citet{kromer2009} using the machinery described by \citet{lucy2002}, and then convert this to a new $r-$packet. For all these processes, the $r-$packet is assumed to retain no information on polarisation and is reemitted in a random direction with zero polarisation\footnote{\citet{hamilton1947} have shown that the angular distribution in resonant line scattering can be expressed as the sum of a dipole and an isotropic term (with relative contribution depending on the quantum numbers of upper and lower state of the line). Although polarisation may arise in cases where the dipole is the dominant term, its magnitude is typically lower than that from electron scattering and thus can be regarded as a second-order effect \citep{jeffery1989,jeffery1991}. }:
\begin{equation}
q_\text{f}^\prime=0   \hspace{1cm} u_\text{f}^\prime=0 ~ .
\end{equation}
If instead the $r-$packet undergoes \textit{electron scattering}, \revised{we follow the scheme introduced by \citet{chandrasekhar1960} and discussed in terms of a Monte Carlo implementation by \citet{code1995} and \citet{whitney2011}}. A scattering angle $\Theta$ is properly sampled (see below), a new direction $\bmath{n_\text{f}^\prime}$ is computed and the Stokes vector is transformed via the scattering matrix 
\begin{equation}
\label{scatt}
A(\Theta) = \frac{3}{4} \begin{pmatrix} $cos$^2\Theta+1 & $cos$^2\Theta-1 & 0 \\ $cos$^2\Theta-1 & $cos$^2\Theta+1 & 0 \\ 0 & 0 & 2\cos\Theta \end{pmatrix} ~.
\end{equation}
After applying the scattering matrix, the dimensionless Stokes vector is normalised so that its first component is equal to $1$. As shown in Fig. \ref{scattering}, since $A(\Theta)$ is defined in the scattering plane, the Stokes vector is first rotated into this plane and then rotated back to the meridian frame after the scattering matrix $A(\Theta)$ has been applied, i.e.
\begin{equation}
\label{dipole} s_\text{f}^\prime = R(\pi-i_2)A(\Theta)R(-i_1)s_\text{i}^\prime ~ ,
\end{equation}
where 
\begin{equation}
R(\phi) = \begin{pmatrix} 1 & 0 & 0 \\ 0 & \cos2\phi & \sin2\phi \\ 0 & -\sin2\phi & \cos2\phi  \end{pmatrix}
\end{equation}
is the matrix to rotate the Stokes vector by an angle $\phi$ clockwise when facing the source. The rotation angles $i_1$ and $i_2$ are computed with the convention that $q^\prime=1$ represents an electric field oscillating in the scattering plane. The scattering angle $\Theta$ is chosen by sampling the probability distribution 
\begin{equation}
\label{probability}
P(\Theta,i_1)= \frac{3}{4} \Big[ ($cos$^2\Theta+1) + ($cos$^2\Theta-1) ( q_\text{i}^\prime\cos2i_1 -  u_\text{i}^\prime\sin2i_1 ) \Big]
\end{equation}
from equation (\ref{dipole}) and using a rejection technique \citep{code1995}.

Finally the Lorentz transformation procedure (see above) with $\,\,\bmath{v}\rightarrow\bmath{-v}$ is used to transform the Stokes vector $S_\text{f}^\prime$ to $S_\text{f}$. 

\subsection{Spectrum extraction techniques}
\label{techniques}

Monte Carlo methods have been exploited in many radiative transfer calculations. These methods have proven particularly useful for multi-dimensional problems since the Monte Carlo algorithm can be easily implemented for arbitrary geometry and scales very well for use on massively parallel compute systems. The main drawback of Monte Carlo methods is their stochastic nature, which leads to solutions that are affected by Monte Carlo noise. Therefore, it is important to attempt to optimise Monte Carlo methods, particularly when we want to extract very weak signals (e.g. polarisation) from the simulations. \revised{In this study we compare} three different methods for extracting synthetic observables from our simulations, with the aim of selecting the most suitable to synthesise spectra with low Monte Carlo noise. In Section \ref{dct} we present a simple direct counting approach, while in Section \ref{ebt} and \ref{tbt} we explore alternative techniques inspired by \citet{lucy1999,lucy2005} and recently implemented by \citet{kerzendorf2014}.

\subsubsection{Direct counting technique}
\label{dct}

In the Direct Counting Technique (DCT), Monte Carlo quanta are followed along their trajectories and their Stokes parameters updated at each interaction, following the procedure outlined in Section \ref{propagation}. Packets that reach the outer boundary are collected in bins according to their final direction $\bmath{n}$ and frequency $\nu$ and the resulting spectra are computed as
\begin{equation}
\begin{pmatrix} I \\ Q \\ U \end{pmatrix} = \sum \frac{\epsilon}{\Delta t~ \Delta\nu~4\pi r^2 }~s_\text{f} ~~,\end{equation}
where $s_\text{f}$ is the dimensionless Stokes vector of the escaping $r-$packet, $\epsilon$ its rf energy, $r$ the distance of the observer from the system and the sum is performed over all the $r-$packets that escaped in the selected angular bin, time interval\break $[t-\Delta t/2,t+\Delta t/2]$ and frequency range $[\nu-\Delta\nu/2,\nu+\Delta\nu/2]$.

The direct counting approach provides a simple way of computing polarisation spectra for different viewing angles, i.e. for different observers. However, the need to average contributions from packets that escape in the same angular bin but with different angles inevitably leads to an approximate result: if the number of angular bins is too small, summing contributions from different angles will produce a poor estimate of the observables seen by a single observer; if instead too many angular bins are used, the number of packets escaping per bin becomes small, leading to high Monte Carlo noise. 

\subsubsection{Event based technique}
\label{ebt}

\begin{figure}
\begin{center}
\includegraphics[width=0.215\textwidth,trim=10pt 47pt 10pt 0pt]{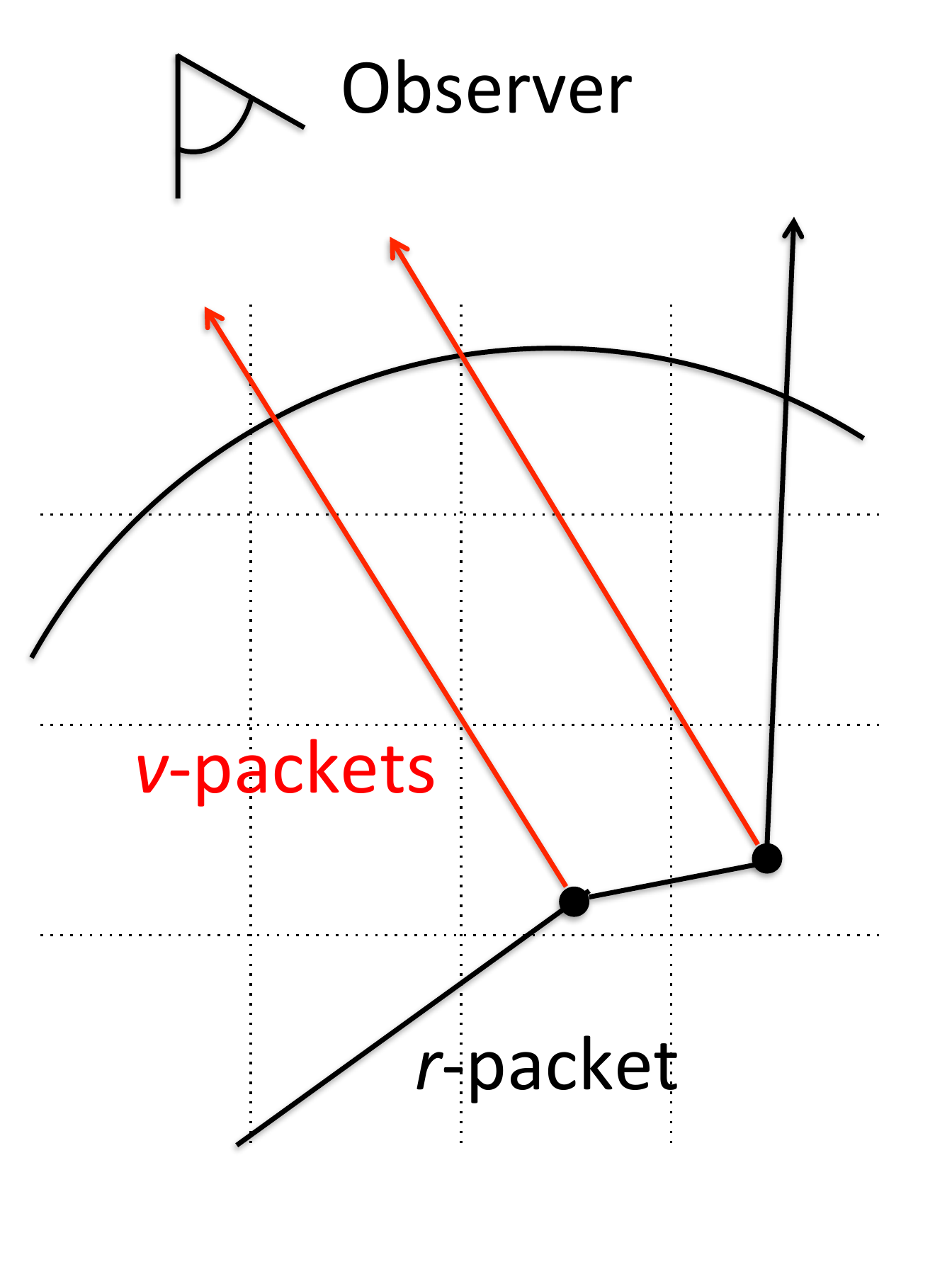}
\includegraphics[width=0.24\textwidth,trim=10pt 47pt 10pt 0pt]{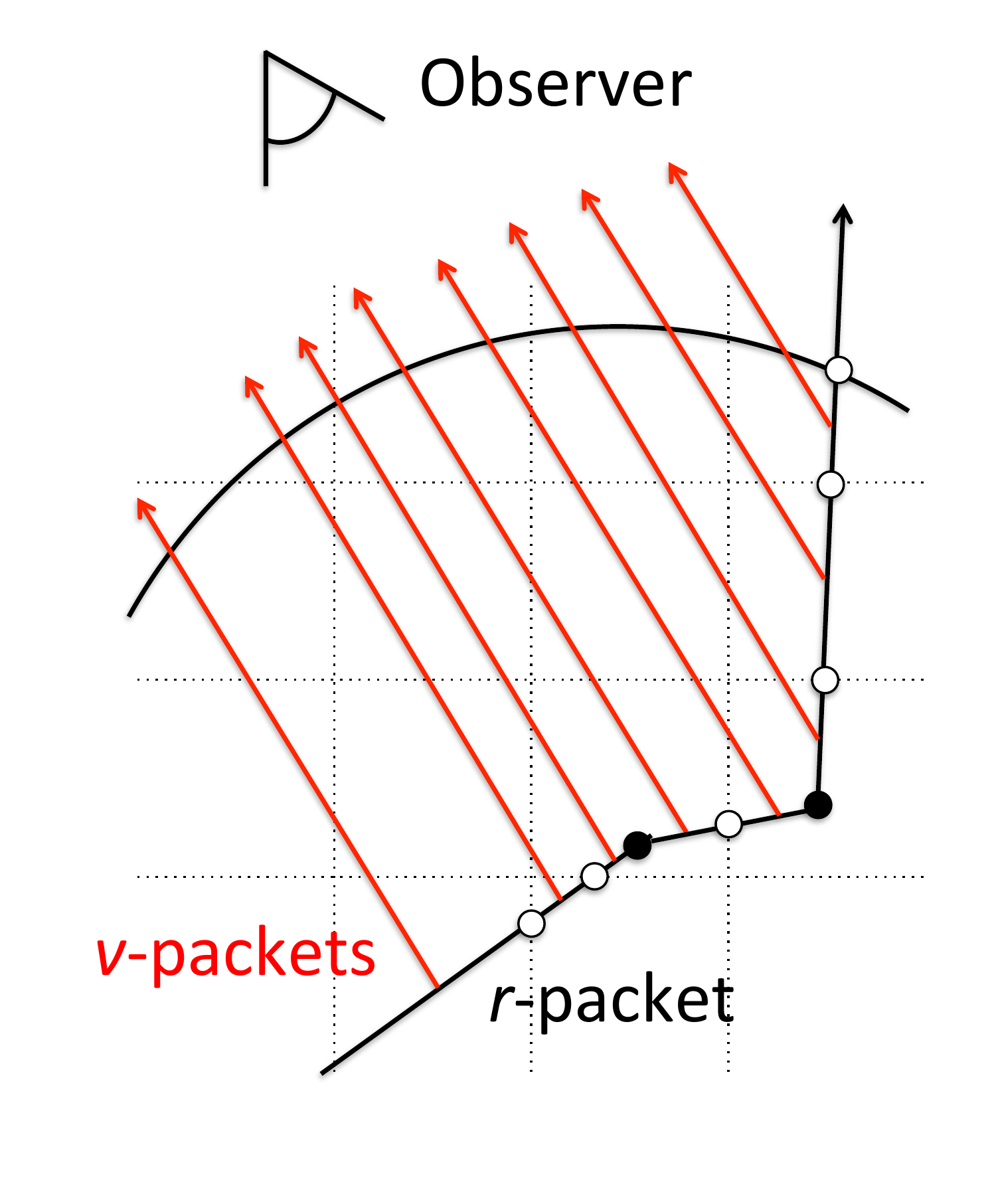}
\caption{Sketches of the principle behind the EBT (left panel) and the TBT (right panel). \revised{EBT}: for every interaction of the $r-$packet, a $v-$packet is created and sent directly to a specific observer. Its Stokes parameters are treated in accordance with the specific interaction and are added as contributions to the emergent spectrum. \revised{TBT}: contributions in the TBT come not only from \revised{trajectories that are terminated by} physical interactions of the $r-$packet (as in the EBT\revised{, black points}) but also from \revised{those terminated by} numerical events ($r-$packets crossing boundaries\revised{, white points}). The arc segment represents the outer boundary of the computational domain. }
\label{virtual}
\end{center}
\end{figure}

In this approach (Event-Based Technique, EBT) we still follow the propagation of Monte Carlo packets exactly as before. However, whenever an $r-$packet interaction occurs the propagation is suspended, and a ``virtual" packet \citep[$v-$packet,][]{kerzendorf2014} is created and handled as described below. Once the $v-$packet calculation is completed, the propagation of the original $r-$packet is resumed and this process repeated for every following interaction. This approach is similar to, and inspired by, that used by \citet{knigge1995} and \citet{long2002}.

When a $v-$packet is created, it is always launched in the direction of a selected observer, $\bmath{n_\text{obs}}$, and has cmf frequency and energy set equal to those of the $r-$packet at the point of creation (see Fig. \ref{virtual}). Since the $v-$packet is forced to go to the observer, the scattering angle is now not determined by sampling either an isotropic distribution ($i-$packet or $k-$packet deactivation) or equation (\ref{probability}) (electron scattering), but rather calculated as
\begin{equation}
\label{costhetaobs}
\cos\Theta= \bmath{n_\text{in}^\prime \cdot n_\text{obs}^\prime} ~.
\end{equation}
The Stokes parameters, initially set equal to those of the incoming $r-$packet, are transformed in accordance with the physical process selected for the $r-$packet. If the $r-$packet is scattered by an electron, the cmf Stokes vector is transformed according to equation (\ref{dipole}), with the scattering angle $\Theta$ determined by equation (\ref{costhetaobs}). If instead a new $r-$packet is created from an $i-$packet or $k-$packet deactivation, we create an unpolarised $v-$packet. 

Once created and assigned a rf frequency, a rf energy ($\epsilon$), a direction of propagation and a Stokes vector ($s_\text{f}$), the $v-$packet is propagated through the ejecta towards the observer and it is interpreted as a contribution to a bin in the emergent spectrum. Specifically, the flux ($I$) and polarisation ($Q$ and $U$) spectra in a given frequency bin $[\nu-\Delta\nu/2,\nu+\Delta\nu/2]$ and time interval $[t-\Delta t/2,t+\Delta t/2]$ can be expressed as
\begin{equation} 
\label{contrib}
\begin{pmatrix} I \\ Q \\ U \end{pmatrix} = \sum \frac{\epsilon}{\Delta t~ \Delta\nu~r^2 }~s_\text{f} \cdot\bigg(\frac{dP}{d\Omega}\bigg|_\text{EBT} ~e^{-\tau_\text{esc}}\bigg)~~,
\end{equation}
where the sum is performed over all the $v-$packets escaping in the selected frequency and time bins, $r$ is the distance of the observer from the system and the two terms inside parentheses are weighting factors accounting for the probability that the $v-$packet could reach the observer. Because the $v-$packet is forced to go to the observer, we first account for the probability per unit solid angle associated with the chosen direction, that is:
\begin{equation} 
\frac{dP}{d\Omega}\bigg|_\text{EBT} = \begin{cases} \frac{1}{4\pi} \hspace{2cm} \text{($k-$/$i-$)packet deact.}  \vspace{0.4cm} \\  \frac{1}{4\pi}P(\Theta,i_1)  \hspace{0.9cm} \text{electron scattering}  \end{cases}~ .
\end{equation} 
The exponential factor in equation (\ref{contrib}) accounts for the probability that the $v-$packet could reach the observer without further interactions, with the optical depth to the boundary, $\tau_\text{esc}$, computed as
\begin{equation} 
\label{tauesc}
\tau_\text{esc} = \tau_\text{cont} + \sum\tau_\text{sob} ~~.
\end{equation}
Here the sum is performed over all the line opacities ($\tau_\text{sob}$) encountered by the $v-$packet on its trajectory to the boundary and the continuum opacity, $\tau_\text{cont}$, is computed as an integral of the continuum attenuation coefficient, $k_\text{cont}$, over trajectory length:
\begin{equation} 
\label{taucont}
\tau_\text{cont} = \int k_\text{cont}~ds ~~.
\end{equation}
\vspace{0.01cm}

The advantage of the $v-$packet technique is that it allows us to compute spectra and light curves for any specific viewing angle, avoiding the need to average contributions from different angles in the same bin. \revised{Unlike in the direct counting approach, where an $r-$packet makes a single contribution to the emergent spectrum, the final spectra in the EBT also contain (appropriately weighted) information derived from every interaction that the $r-$packet undergoes (see Fig. \ref{virtual})}. For these reasons, we expect this technique to produce spectra and light curves with lower Monte Carlo noise. However, the need of creating and handling $v-$packets introduces a computational overhead that could make the EBT less efficient than the DCT. Such possibilities are explored quantitatively in Section \ref{comparetech}. 

\subsubsection{Trajectory based technique}
\label{tbt}

\revised{In the EBT described above, we used all the $r-$packets to provide us with an ensemble of physical events, and then for each event in the ensemble we computed the probability of it giving rise to a photon escaping to the observer. In the third technique (Trajectory-Based Technique, TBT) we instead obtain from the $r-$packets an ensemble of photon trajectories, which can be taken as a discrete representation of the radiation field in the simulation. We can then estimate observables by summing over this ensemble of trajectories, computing for each the probability that interactions of radiation on that path could have given rise to photons escaping to the observer.} 

\revised{This summation is achieved by generating a $v-$packet for each trajectory path $\Delta l$ (including those terminated by numerical events, e.g. packets crossing grid cell boundaries) and sending it towards the observer at $\bmath{n_\text{obs}}$\break (see Fig. \ref{virtual}). The $v-$packet contribution to the emergent spectrum is first weighted by the probability per unit solid angle ($dP/d\Omega|_{TBT}$) that photons on the path $\Delta l$ could have undergone an interaction that gave rise to re-emission/scattering towards the observer.  As with the EBT, we must also account for the probability of any scattered/emitted radiation reaching the observer via a suitable mean exponential factor, $< e^{-\tau_\text{esc}} >$. Thus, in the TBT, we compute synthetic observables via}
\begin{equation} 
\begin{pmatrix} I \\ Q \\ U \end{pmatrix} = \sum \frac{\epsilon}{\Delta t~ \Delta\nu~r^2 }~s_\text{f} \cdot\bigg(\frac{dP}{d\Omega}\bigg|_\text{TBT} ~ < e^{-\tau_\text{esc}} >\bigg)~,
\end{equation}
\revised{where the summation is again over $v-$packets in the selected frequency and time bins. Formally, the exponential weight factor of a $v-$packet in the TBT should be computed as}
\begin{equation} 
< e^{-\tau_\text{esc}} > = \frac{\int\limits^{l+\Delta l }_{l} e^{-\tau_\text{esc}(l^\prime)}~dl^\prime}{\Delta l} ~~,
\end{equation}
\revised{where the integral runs over the trajectory path $\Delta l$. To first order, however, this can be approximated by generating the $v-$packet at the mid point of $\Delta l$ and computing the exponential factor from its flight (as described in Section \ref{ebt}); i.e.}
\begin{equation} 
< e^{-\tau_\text{esc}} > = e^{-\tau_\text{esc}(l+\Delta l / 2)} ~.
\label{meanexp}
\end{equation}
\revised{In principle, $dP/d\Omega|_{TBT}$ can be formulated to account for all (effective) scattering/fluorescence processes. However, since we are primarily interested in studying contributions to the emergent polarisation spectrum, we focus only on electron scattering, for which}
\begin{equation} 
\frac{dP}{d\Omega}\bigg|_\text{TBT} = \frac{1}{4\pi}~P(\Theta,i_1) ~ k_\text{sc}\Delta l ~ ,
\end{equation}
\revised{where $k_\text{sc}$ is the scattering attenuation coefficient.}

A key difference between the EBT and the TBT is that, in the latter, every trajectory element of the $r-$packets contributes to the synthetic observables, whereas, in the former, only physical interaction events contribute. For instance, in the limit of optically thin ejecta many more $v-$packets would contribute to the emergent spectrum in the TBT compared to the EBT (see Fig. \ref{virtual}). However, a drawback of the TBT is that \revised{$\tau_\text{esc}(l + \Delta l/2)$} should describe the mean probability of escape for points along the trajectory element \revised{(see equation \ref{meanexp})}, rather than the exact probability of escape from the interaction point, as in the EBT. For an $r-$packet trajectory with moderate optical depth ($\tau\gtrsim1$), this approximation may lead to a poor estimate of the observables. Breaking the trajectory into smaller paths (with $\Delta\tau\ll1$) and generating $v-$packets at each midpoint may be required, slowing down the code and making the TBT less efficient then the EBT (see Section \ref{comparetech}).

\begin{figure}
\includegraphics[width=0.47\textwidth,trim=0pt 15pt 0pt 0pt]{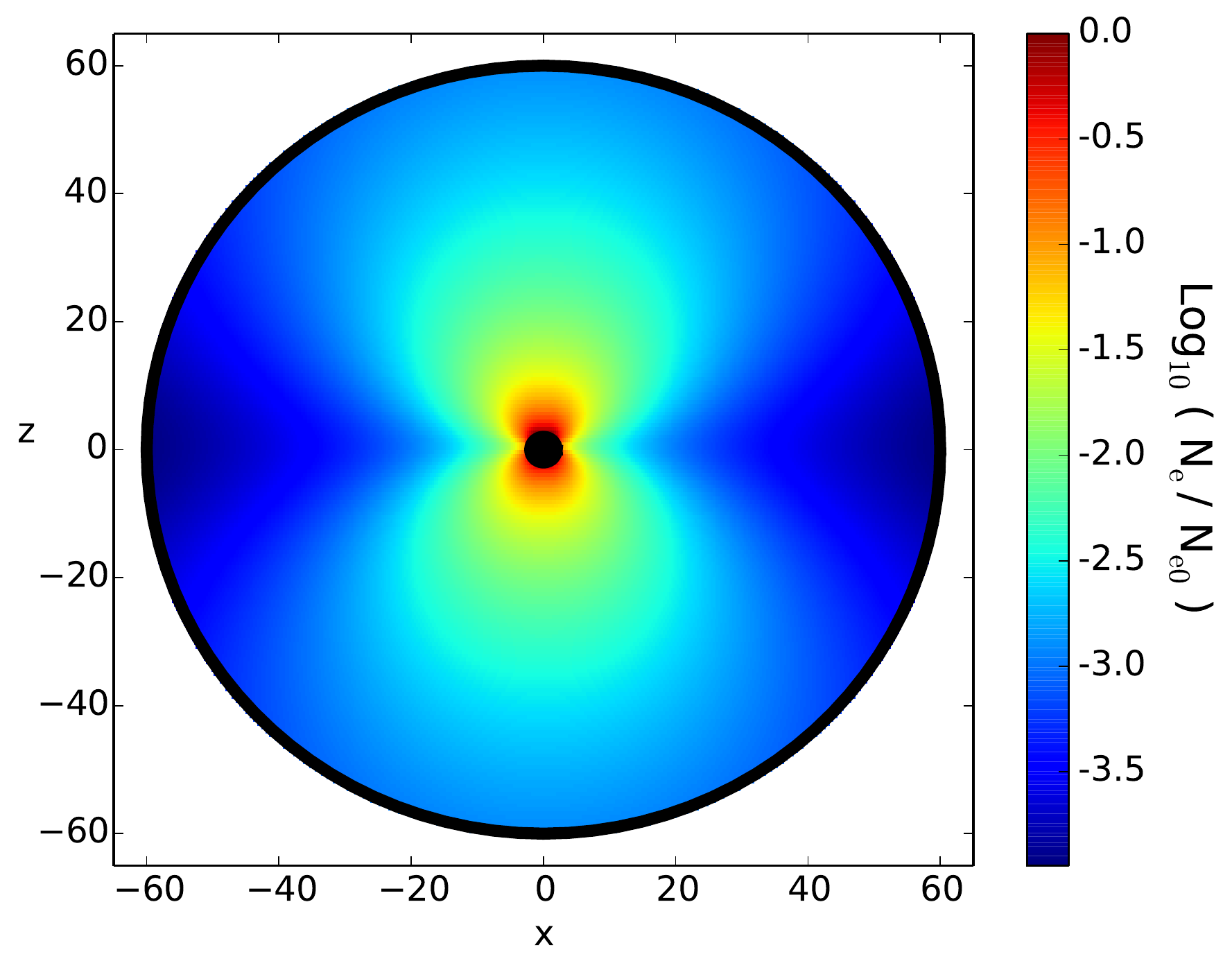}
\centering
\caption{The geometry adopted by \citet{hillier1994} and used for our test calculations. Monte Carlo quanta are created inside a spherical shell of radius $R_\text{min}$, propagate into a region with a prolate density distribution and are free to escape when they reach the outer spherical shell at $R_\text{max}=30.0R_\text{min}$. $N_\text{e}(r,\beta)$ is the electron density distribution and $N_\text{e0}=N_\text{e}(R_\text{min},0)$. }
\label{hillier}
\end{figure}

\section{Test code}
\label{testcode}

In the following we present a simple test code, with the aim of validating our polarisation scheme (Section \ref{comparehillier}) and selecting the most suitable of the techniques described in Section \ref{techniques} to synthesise spectra with low Monte Carlo noise (Section \ref{comparetech}). In this code, packets are generated with null polarisation in a small inner sphere (to mimic a point source) and then allowed to propagate into an envelope where either interactions with electrons or continuum absorptions can occur. Here time evolution for the ejecta is neglected.

\subsection{Polarisation scheme validation}
\label{comparehillier}

\begin{figure}
\includegraphics[width=0.46\textwidth]{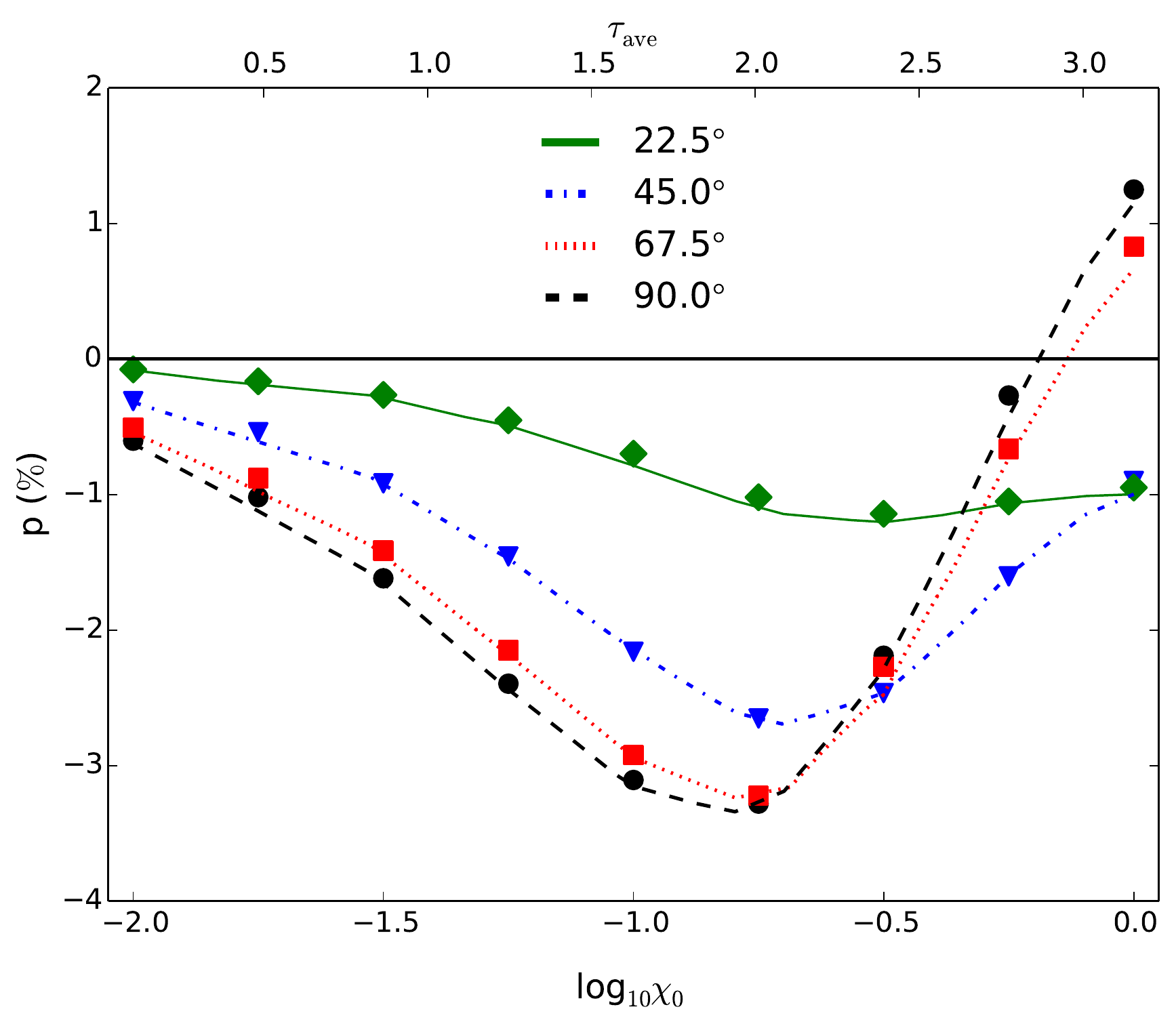}
\centering
\caption{Continuum polarisation as a function of $\chi_0$ and $\tau_\text{ave}$ for 4 different viewing angles for the setup described in Section \ref{comparehillier}. Symbols indicate predictions from our test calculations, while the lines are reported from \citet{hillier1994} for comparison. \revised{The Monte Carlo noise error bars are not shown since they are smaller than the symbol sizes.}}
\label{hillier1}
\end{figure}

\begin{figure}
\includegraphics[width=0.47\textwidth,trim=0pt 0pt 0pt -23pt]{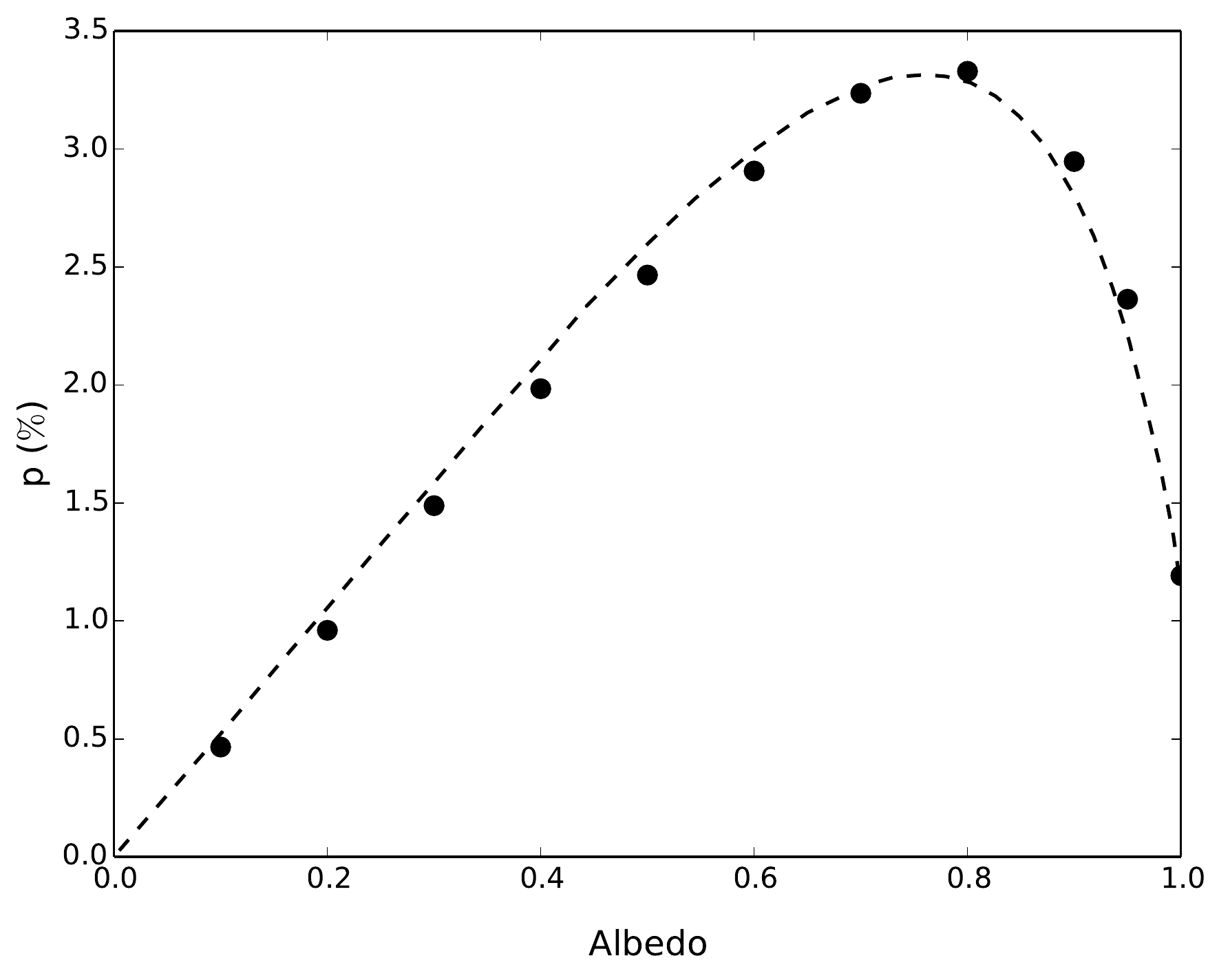}
\centering
\caption{Continuum polarisation as a function of the albedo (\revised{black} points) for the setup described in Section \ref{comparehillier}, together with the predicted curve (dashed line) from \citet{hillier1994}. \revised{Here the observer's inclination is $90^\circ$ and $\chi_0=0$. The Monte Carlo noise error bars are not shown since they are smaller than the symbol sizes.}}
\label{hillier2}
\end{figure}

To validate the polarisation scheme, we first focus our attention on the DCT and choose to reproduce a simple configuration described by \citet{hillier1994}. As shown in Fig. \ref{hillier}, a point source is surrounded by a detached spherical shell with inner radius $R_\text{min}=2.0$ and outer radius  $R_\text{max}=30.0R_\text{min}$, with a prolate density distribution $N_\text{e}(r,\beta)$ such that
\begin{equation}
\sigma_\text{e}N_\text{e}(r,\beta) = \chi_0 \bigg( \frac{R_\text{min}}{r} \bigg)^2 (1+10~$cos$^2\beta) \hspace{0.3cm},
\end{equation}
where $\sigma_\text{e}$ is the Thomson cross section and $r$ and $\beta$ express the radius and the polar angle inside the envelope. The $\chi_0$ parameter is related to the solid-angle averaged (from inner to outer boundary) optical depth, $\tau_\text{ave}=2.888\chi_0$, and can be varied to investigate the impact of different scattering optical depth on the polarisation signal. Neglecting absorption and assuming a pure electron scattering envelope, the continuum polarisation as a function of $\chi_0$ is shown in Fig. \ref{hillier1} for four different viewing angles $i$ ($22.5^\circ$, $45^\circ$, $67.5^\circ$ and $90^\circ$). The agreement between our predicted values and the expected curves from \citet{hillier1994} is encouraging. We also carried out calculations in which we include continuum absorption opacity. These show good agreement with the predicted dependence of the continuum polarisation on the albedo (ratio of the scattering to the total opacity, see Fig. \ref{hillier2}). 

\subsection{Comparison between different techniques}
\label{comparetech}

\begin{figure*}
\includegraphics[width=0.47\textwidth]{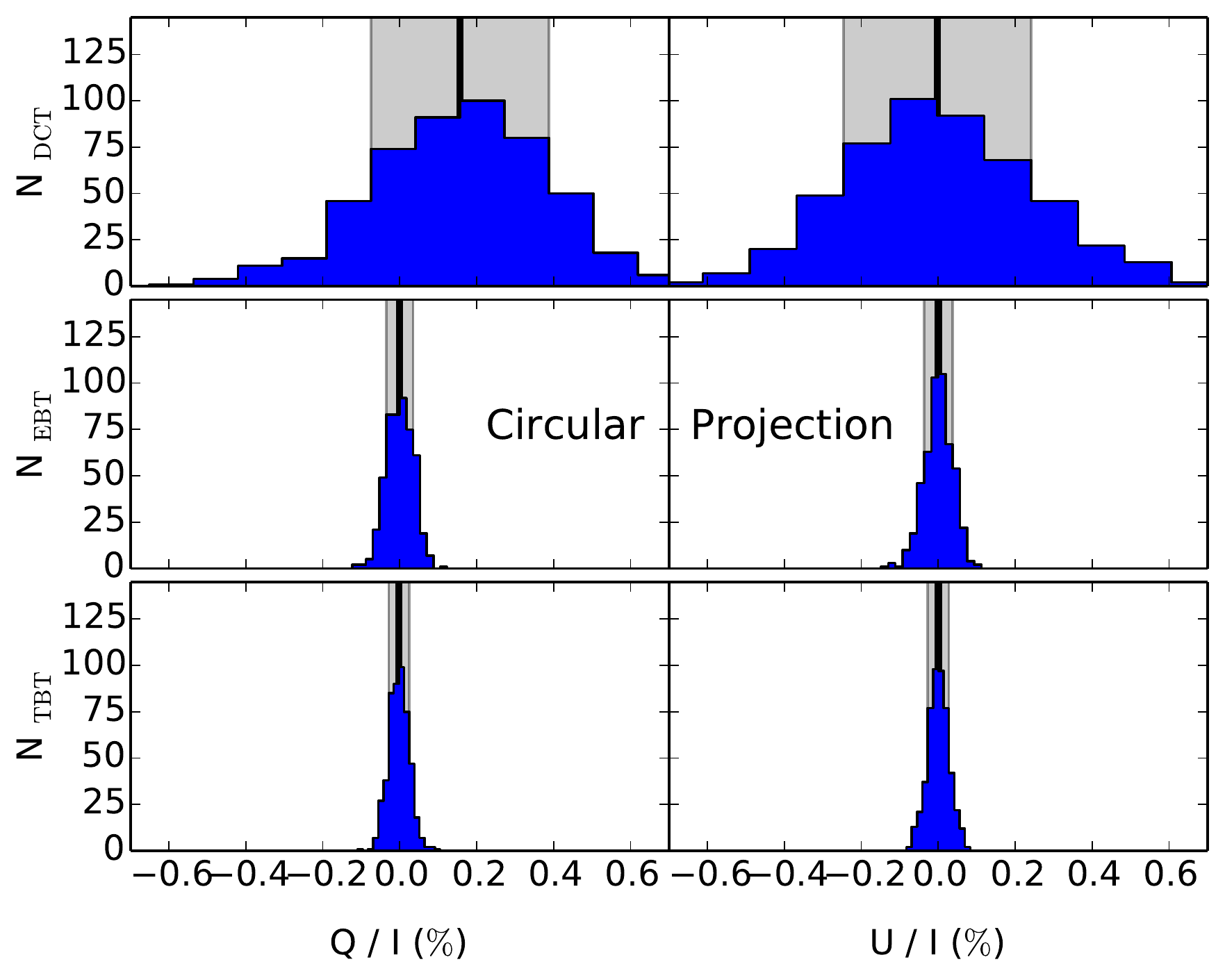}
\includegraphics[width=0.47\textwidth]{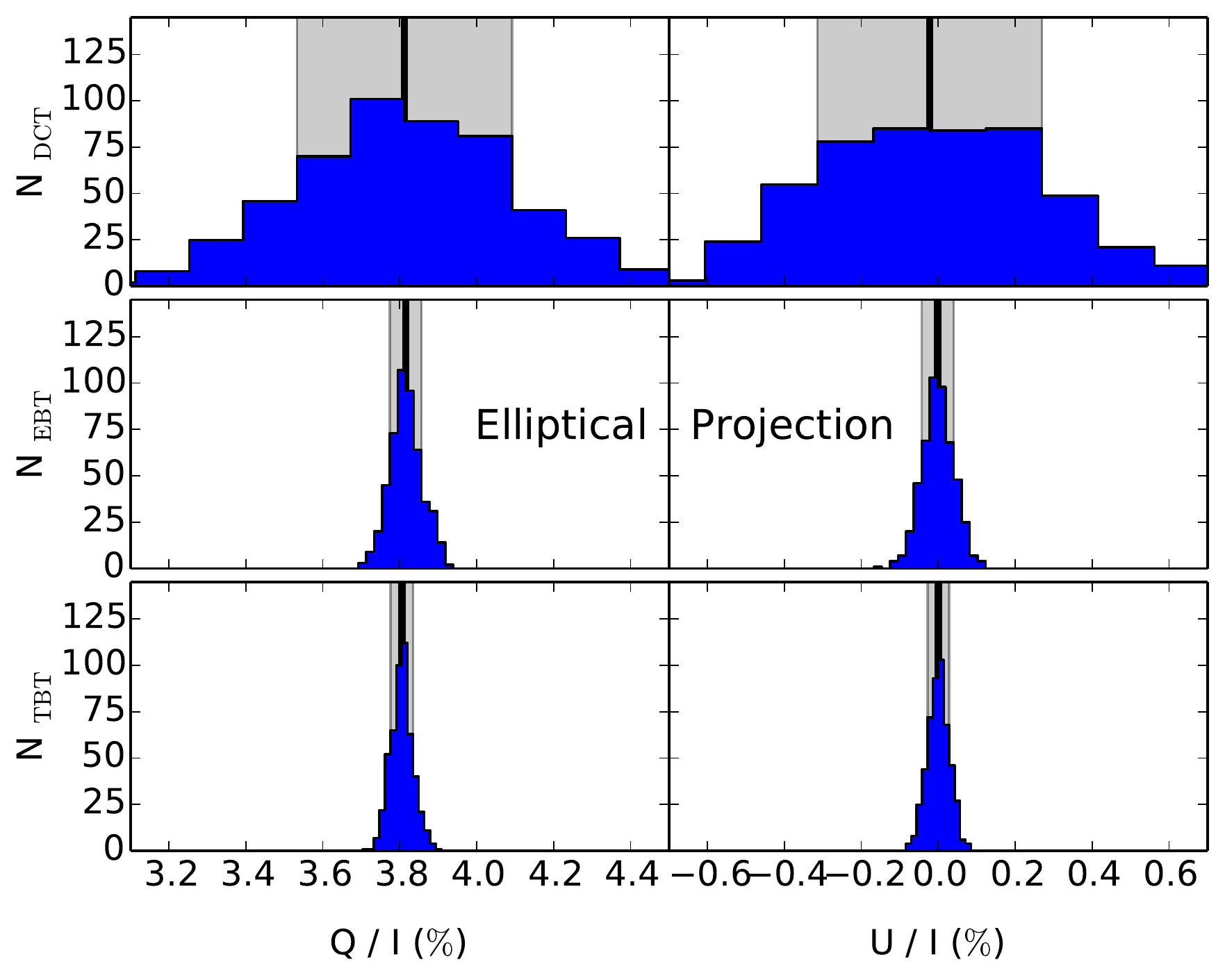}
\caption{$Q$ and $U$ continuum polarisation distributions of $N_{sim}=500$ runs with the DCT (top panels), the EBT (middle panels) and the TBT (lower panels). The adopted geometry is an oblate ellipsoid with axis ratio of two, as described in Section \ref{comparetech}. The system is viewed along the minor axis (circular projection, left panel) and the major axis (elliptical projection, right panel). \revised{For each distribution, a solid vertical line indicates the average value, $\bar{x}$, and the grey shaded area marks $\pm$ one standard deviation, $\sigma$. The average values and the standard deviations of each plot are reported in Table \ref{tab1} together with the average runtimes, $\bar{t}$.}}
\label{comparison}
\end{figure*}

A convenient means to compare the three techniques for extracting observables outlined in Section \ref{techniques} is by studying their accuracy in reproducing continuum polarisation for a given configuration. We did this by repeatedly running our test code a number ($N_\text{sim}=500$) of times \revised{for each technique (with different random number seeds determined by the wall-clock time)} and comparing the distributions of polarisation values obtained using each method. For these experiments, we chose a configuration in which a point source illuminates a surrounding atmosphere, chosen to be a constant density oblate ellipsoid with axis ratio of two, i.e.
\begin{equation}
\frac{x^2}{a^2}+\frac{y^2}{b^2}+\frac{z^2}{c^2}=1 ~~, \hspace{1cm}  a=b=2c ~~ .
\end{equation}
In all the simulations, $10^6$ packets have been created, a pure scattering atmosphere has been assumed (i.e. no line opacity and albedo equal to 1) and the scattering coefficient has been set to $k_\text{sc}=1 / a$ and kept constant throughout the ejecta. As mentioned in Section \ref{dct}, continuum polarisation levels estimated from the DCT may be either inaccurate or \revised{have large} uncertainties depending on whether the number of angular bins is small or \revised{large}, respectively. We found that a value of $51$ angular bins provides a reasonable compromise and we therefore adopt this in the DCT calculations. The results of the comparison between the three techniques are shown in Fig. \ref{comparison} and reported in Table \ref{tab1}. 

For an observer along the $y$-axis (Fig. \ref{comparison}, left panel), the projection of the model along the line-of-sight is circular and we expect to find null polarisation (see above). The $Q$ and $U$ values in every technique are indeed consistent with zero but, because of Monte Carlo noise, a distribution of values is obtained. The width of this distribution provides a convenient quantification of the Monte Carlo noise introduced by each method. As expected, the standard deviations obtained with the EBT and TBT methods are smaller than with the DCT method, by factors of $\sim6.6$ and $\sim8.6$, respectively. Given that Monte Carlo noise is expected to scale with the square root of the number of packets, the DCT could reach the same signal-to-noise ratio as the EBT (TBT) with a factor of $\sim45$ ($\sim75$) more packets. Although the $v-$packet routine introduces computational overheads in the EBT and TBT (see Table \ref{tab1}), the direct counting approach would still be a factor of $\sim7.5$ ($\sim2.5$) slower than the $v-$packets technique. We also note that, even with the same signal-to-noise ratio, the DCT would be less accurate in predicting polarisation values because of the need to average contributions from different viewing angles (via angular binning): indeed, closer inspection shows that the $Q$ distribution for the DCT is shifted towards positive values (in the binning approach contributions to the spectra come from $r-$packets escaping close to, but not exactly along, the $z$ direction and thus the average value of $Q$ is slightly positive). 

\begin{table}
\centering
\caption{Average values and standard deviations of the distributions of $Q$ and $U$ values predicted on $500$ simulations by the DCT, EBT and TBT. The system is an oblate ellipsoid with axis ratio of two viewed along the $z$-axis (circular projection) and along the $x$-axis (elliptical projection). The averaged runtime, $\bar{t}$, is reported for each distribution.}
\label{tab1}
\begin{tabular}{lccc}
\hline
Circular & $\bar{Q}\pm\sigma_Q$ (per cent) & $\bar{U}\pm\sigma_U$ (per cent) & $\bar{t}$ (s) \\
\hline
DCT & 0.16 $\pm$ 0.23 & 0.00 $\pm$ 0.24 & 3.9 \\
EBT & 0.00 $\pm$ 0.04 & 0.00 $\pm$ 0.04 & 22.3 \\
TBT & 0.00 $\pm$ 0.03 & 0.00 $\pm$ 0.03 & 123.0 \\
\hline
Elliptical & $\bar{Q}\pm\sigma_Q$ (per cent) & $\bar{U}\pm\sigma_U$ (per cent) & $\bar{t}$ (s) \\
\hline
DCT & 3.81 $\pm$ 0.28 & -0.02 $\pm$ 0.29 & 3.9 \\
EBT & 3.82 $\pm$ 0.04 & 0.00 $\pm$ 0.04 & 26.3 \\
TBT & 3.81 $\pm$ 0.03 & 0.00 $\pm$ 0.03 & 144.5 \\
\hline
\end{tabular}
\end{table}

If viewed down the $x$-axis (right panel), the projection becomes an ellipse and thus we expect to see a polarisation signal. The three techniques agree in reproducing a continuum polarisation of $p\sim3.8$~per cent but, again, the DCT gives a much broader distribution of values, indicating that it is more severely affected by noise in the simulation. In order to reach the same standard error of the mean of the EBT (TBT), the DCT would require a factor of $\sim50$ ($\sim100$) more packets, making the total runtime a factor of $\sim7.5$ ($\sim3$) longer than the $v-$packets scheme. 

This simple comparison shows that the $v-$packet approaches are more precise in estimating polarisation, allowing us to reach a given Monte Carlo noise with many fewer Monte Carlo quanta (and substantially shorter runtimes) than the simple DCT would require. As already anticipated in Section \ref{tbt}, the EBT is indeed more efficient than the TBT because the runtime for the latter is limited by the need of breaking $r-$packet trajectories with moderate optical depth ($\tau\gtrsim1$) into smaller paths (with $\Delta\tau\ll1$), in order to give accurate results for $\tau_\text{esc}$. We note that, although we have carried out polarisation tests here, this improvement in Monte Carlo noise could also be exploited for extracting high-quality observables of any sort.

\section{ARTIS} 
\label{artis}

In this Section we describe the implementation of our polarisation scheme into the three-dimensional, time-dependent radiative transfer code ARTIS \citep{kromer2009} and test it for one-dimensional and two-dimensional models. Section \ref{implemartis} outlines the implementation of the different techniques described in Section \ref{techniques}. In Section \ref{w7} we test the code using the one-dimensional W7 explosion model \citep{nomoto1984,iwamoto1999} and check the accuracy of the $v-$packet technique in computing spectra and reproducing continuum polarisation consistent with zero. Finally, in Section \ref{ellipsoidal} we apply the new version of the code to two-dimensional ellipsoidal models to investigate the impact of simple aspherical geometries on line and continuum polarisation for different viewing angles, and compare our results to those of similar studies made using other codes.

\subsection{Implementation}
\label{implemartis}

Polarisation is implemented into ARTIS by assigning a Stokes vector to each $r-$packet and by transforming this according to the physical process the packet undergoes (see Section \ref{propagation}). For the DCT, the same binning approach already used in the code for spectra and light curves is extended to compute polarisation spectra. 

As mentioned in Section \ref{tbt}, the $v-$packet TBT should yield spectra with lower Monte Carlo noise compared to the EBT since contributions to the polarisation spectra come from every event in the $r-$packet histories, including both physical interactions and numerical events (e.g. packets crossing grid cell boundaries). However, accurate results from the TBT require that care is taken in the calculation of $\tau_\text{esc}$, which can introduce a large computational overhead for complicated opacity distributions. Indeed, our test calculations (Section \ref{comparetech}), suggested that this additional computational overhead can ultimately make the TBT less efficient than \revised{the} EBT. Consequently, here we have chosen to implement a $v-$packet routine using the EBT that can be used to compute synthetic observables from ARTIS.

The $v-$packet routine allows us to compute flux and polarisation spectra for multiple observers, simply by using a loop to generate $v-$packets over a set of different viewing angles. Several input parameters can be chosen to optimise performance in the calculations. First, the calculation of the optical depth $\tau_\text{esc}$, see equation (\ref{tauesc}), can be stopped when the $v-$packet reaches a maximum value $\tau_\text{esc}^\text{max}$: $v-$packets with high optical depth to the boundary would make vanishingly small contributions to the final spectrum (because of the exponential factor, see equation \ref{contrib}) and can thus be neglected. Since the computation cost of the $v-$packet methods is dominated by the calculation of $\tau_\text{esc}$, the runtime is strongly affected by this parameter. We have carried out test calculations to verify that setting $\tau^\text{max}_\text{esc}=10$ and neglecting $v-$packets with higher optical depth does not affect the final result, and adopted this as our default value for all the calculations presented here. Our implementation also includes the option to generate $v-$packets only in a selected spectral interval. Because spectropolarimetric observations usually cover the optical region of the spectrum, our default wavelength range is $3500-10\,000$~\AA. Given that much of the runtime of the code can be consumed in computing $\tau_\text{esc}$ for packets in the bluer regions, which easily reach $\tau_\text{esc}^\text{max}$ because of strong iron-line blanketing, this particular cut in wavelength speeds up the calculations by a factor of $\sim4$ compared to calculations with no wavelength cut. Finally, the $v-$packet routine can be switched on or off for timesteps as chosen by the user (note that the activation or deactivation of $v-$packets has no effect on the $r-$packet propagation). 

\subsection{W7 model}
\label{w7}

\begin{figure}
\includegraphics[width=0.44\textwidth,trim=20pt 0pt 0pt 0pt]{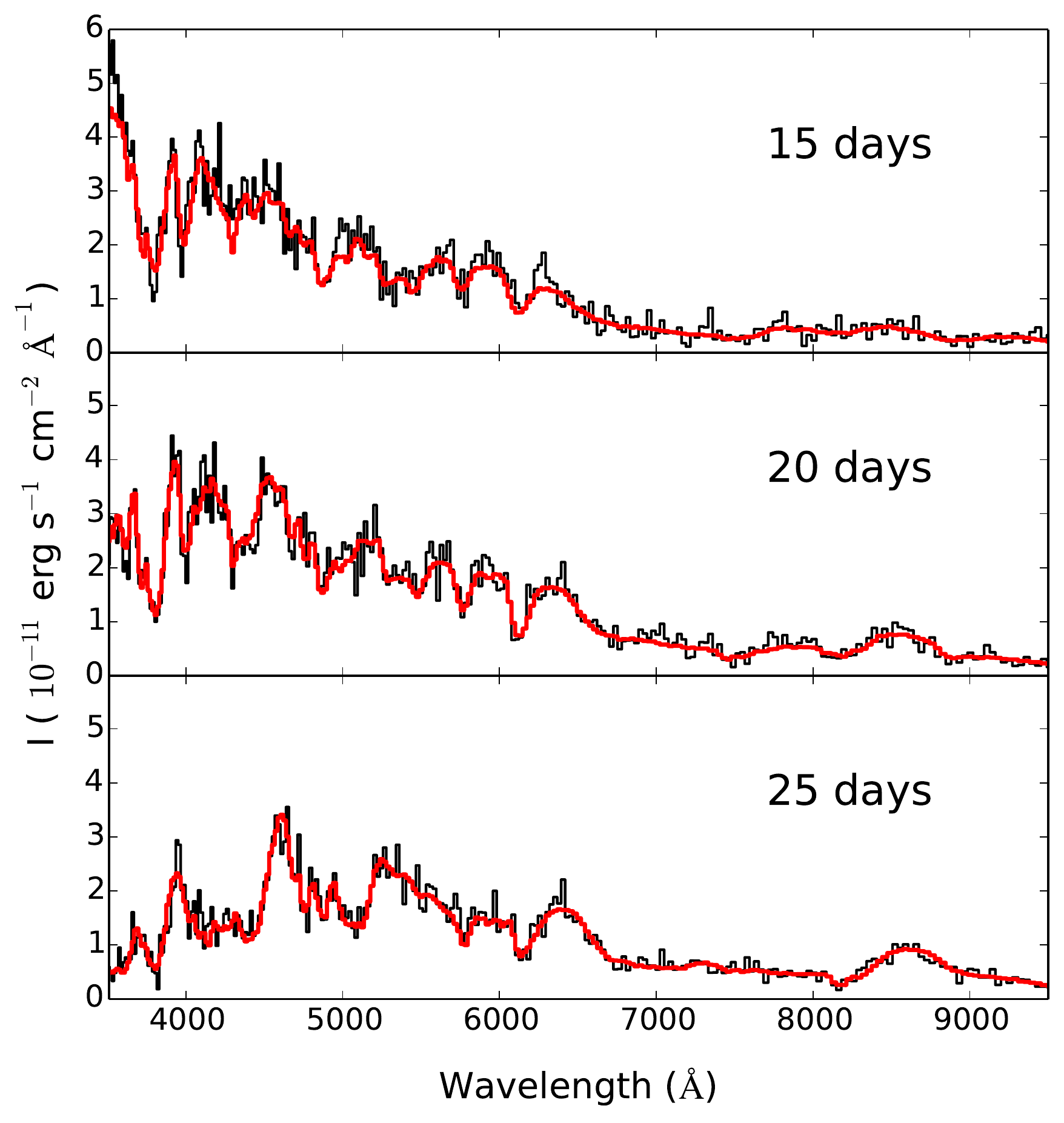}
\centering
\caption{\revised{Spectra for the W7 model at $15$, $20$ and $25$ days after explosion computed with the DCT (black lines) and EBT (red lines). The spectra are computed for an observer at\break $\bmath{n_\text{obs}}=(0,0,1)$. The model supernova is assumed to be at $1$~Mpc.}}
\label{W7spectra}
\end{figure}

Although much effort has been recently directed at developing multi-dimensional explosion models \citep{rosswog2009,jordan2012b,kushnir2013,moll2014,fink2014}, the one-dimensional parameterised deflagration model W7 is still widely used since its composition and structure provide reasonable agreement with observations of (``normal") SNe Ia \citep{kasen2006,kromer2009,jack2011,vanrossum2012,gall2012}. Here we calculate ARTIS flux and polarisation spectra for this model aiming to: (i) compare the accuracy of the DCT and the EBT in reproducing flux spectra at different epochs; (ii) test our polarisation implementation on a spherically symmetric system for which null polarisation spectra are expected.

For this calculation we simulate $8\times10^7$ Monte Carlo packets and compute spectra over $111$ logarithmic time-steps from $2$ to $120$ days after explosion. \revised{We bin the final spectra of both the DCT and the EBT in logarithmic wavelength bins with $\frac{\Delta\lambda}{\lambda} = 3.912\times10^{-3}$}. For this test, local thermodynamic equilibrium (LTE) has been assumed for all time-steps\footnote{We note that LTE is a crude approximation, especially for epochs after maximum light. We will, however, confine most of our discussion to relatively early epochs when the LTE approximation should be reasonable.}. The calculation was carried out by mapping the spherically symmetric W7 model onto a $100^3$ Cartesian grid, through which the packets were propagated. The $v-$packet EBT is activated from $10$ to $30$ days after explosion and only for $r-$packets with emergent rf wavelength between $3500$ and $10\,000$~\AA. Spectra for the EBT are computed for the viewing angle $\bmath{n_\text{obs}}=(0,0,1)$, although we note that (since the model is spherically symmetric) the choice of observer orientation here is arbitrary. Compared to the DCT, the runtime penalty associated with using the $v-$packet routine in the EBT is found to be less than a factor of two, with the advantage that the number of packets contributing to the emergent spectrum is a factor of $\sim115$ higher.

\begin{figure}
\includegraphics[width=0.44\textwidth,trim=20pt 0pt 0pt -7pt]{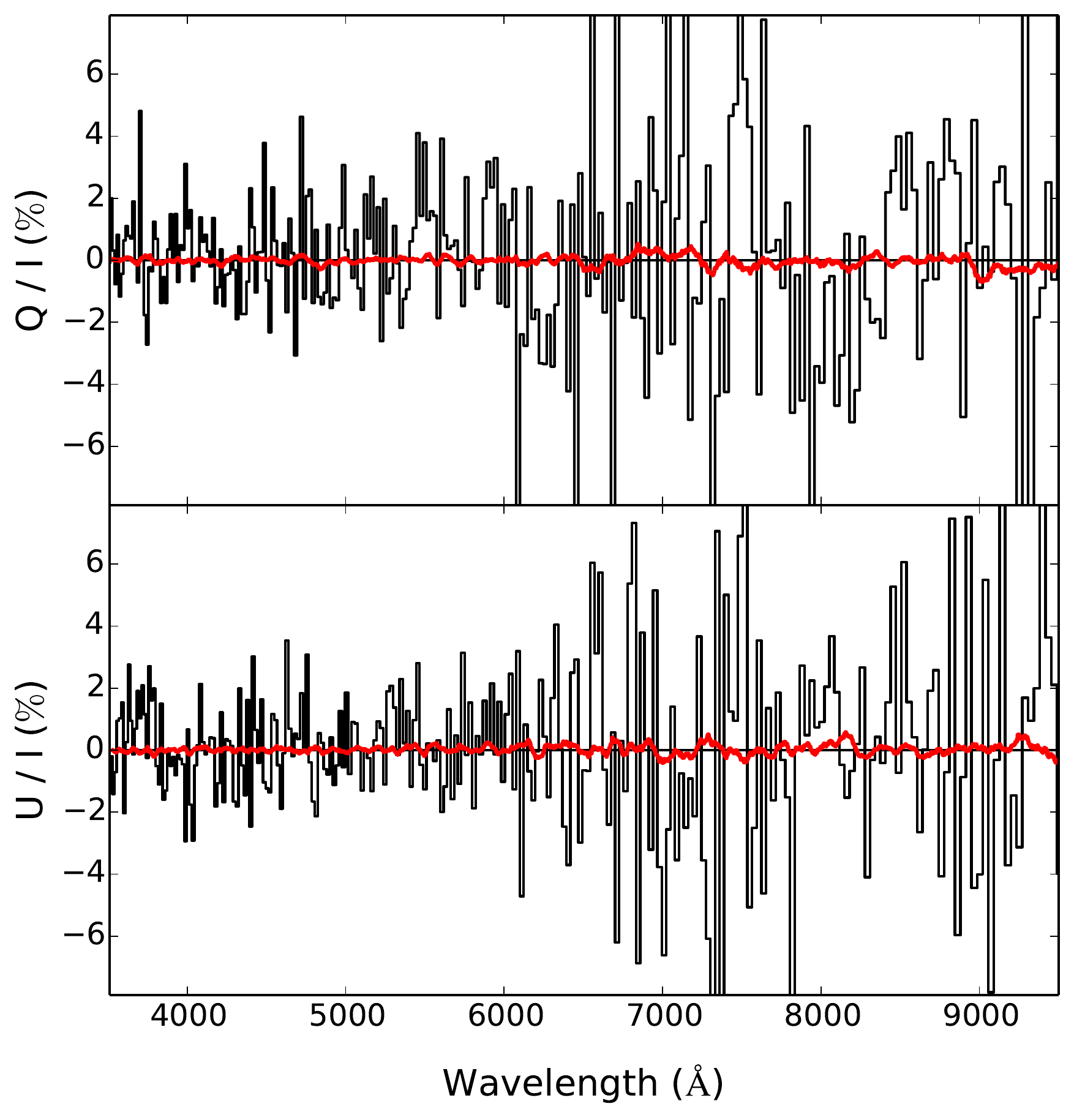}
\centering
\caption{Accuracy of the DCT and the EBT in reproducing continuum polarisation consistent with zero. The W7 model has been used for this test calculation. Polarisation spectra are computed at $20$ days after explosion with the DCT (black lines) and EBT (red lines). The increase in Monte Carlo noise at longer wavelengths is due to the lower flux in the spectrum (see Fig. \ref{W7spectra}). }
\label{W7pol}
\end{figure}

Fig. \ref{W7spectra} shows spectra calculated with the EBT at $15$, $20$ (around $B$ band maximum light) and $25$ days after explosion. Angle-resolved ($10$ angle bins\footnote{\revised{In Section \ref{comparetech} we chose a number of 51 viewing-angle bins as a reasonable value to obtain accurate angle-resolved results with relatively low Monte Carlo noise for a simple ellipsoidal configuration. However, the number of packets used here requires a smaller number of bins in order to achieve a reasonable level of Monte Carlo noise in the spectra. Reducing the number of viewing-angle bins to 10 does not affect the accuracy of the results (given that the observables in a 1D model are the same for different viewing angles) but instead merely decreases the Monte Carlo noise in the predicted spectra.}}) spectra obtained with the DCT are shown for comparison. We note that the calculation of the direct counting flux spectrum is exactly the same as in the previous version of ARTIS \citep{kromer2009}, with the exception that electron scattering is now treated fully via the scattering matrix in equation (\ref{scatt}) rather than assuming isotropic scattering. The agreement between the two techniques is very good, with the EBT being much less affected by Monte Carlo noise, as expected. To estimate the Monte Carlo noise in the spectra, we use the fact that the calculation has been carried out \revised{on multiple cores which provides us with a set of independent estimates for any given observable. In particular, we divide the simulation outputs into eight subsets, each comprising one eighth of the cores, and calculate an emerging spectrum for each of them. Spectral differences between different subsets are} representative of the Monte Carlo noise and estimated by computing residuals from a mean spectrum. The standard deviation of the residual in the $v-$packet \revised{spectrum is $13.3$ times} smaller than that calculated for the angle-resolved direct counting spectrum.

\begin{figure}
\includegraphics[width=0.46\textwidth]{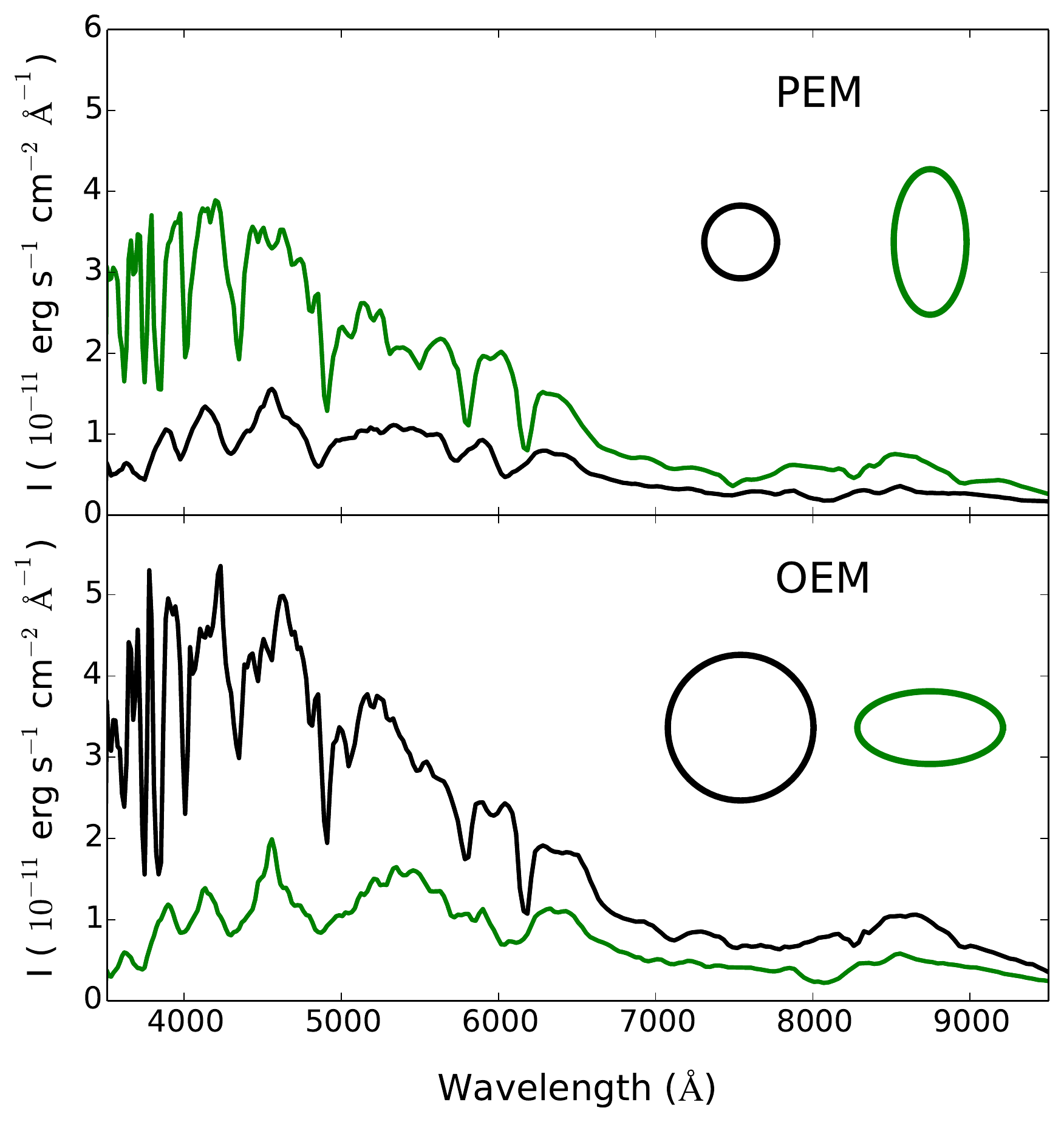}
\centering
\caption{Spectra for the PEM (top panel) and the OEM (bottom panel) calculated with the EBT at $19$ days after explosion. Black/green lines are for an observer orientation along $z$/$x$ ($\bmath{n_\text{obs,1}}$/ $\bmath{n_\text{obs,2}}$ in the text). Scaled projected surfaces are shown for each viewing angle. }
\label{ellipse_spectra}
\end{figure}

Polarisation spectra around maximum light in the $B$ band are reported in Fig. \ref{W7pol}. As expected from a one-dimensional model, the average $Q$ and $U$ throughout the whole spectral range are consistent with zero for both techniques, with the signal-to-noise decreasing towards the red because of the lower flux level. The decrease in Monte Carlo noise when comparing the DCT to the EBT is remarkable: \revised{the standard deviation in the $Q$ ($U$) spectrum is a factor of $14.1$ ($13.7$) larger} in the former compared to the latter, in good agreement with our findings from the flux spectra. This simple comparison clearly shows that the $v-$packet technique is superior for producing accurate polarisation levels. 

\revised{We note that the factor by which the noise improves does depend on the number of angular bins used (the improvement is less dramatic - but still significant - if fewer bins are used). However, our choice of 10 bins is rather conservative (c.f. Section \ref{comparetech} where it was found that $\sim51$ bins were required for accurate representation of a simple 2D model).}

\subsection{Ellipsoidal toy model}
\label{ellipsoidal}

\begin{figure*}
\includegraphics[width=0.43\textwidth]{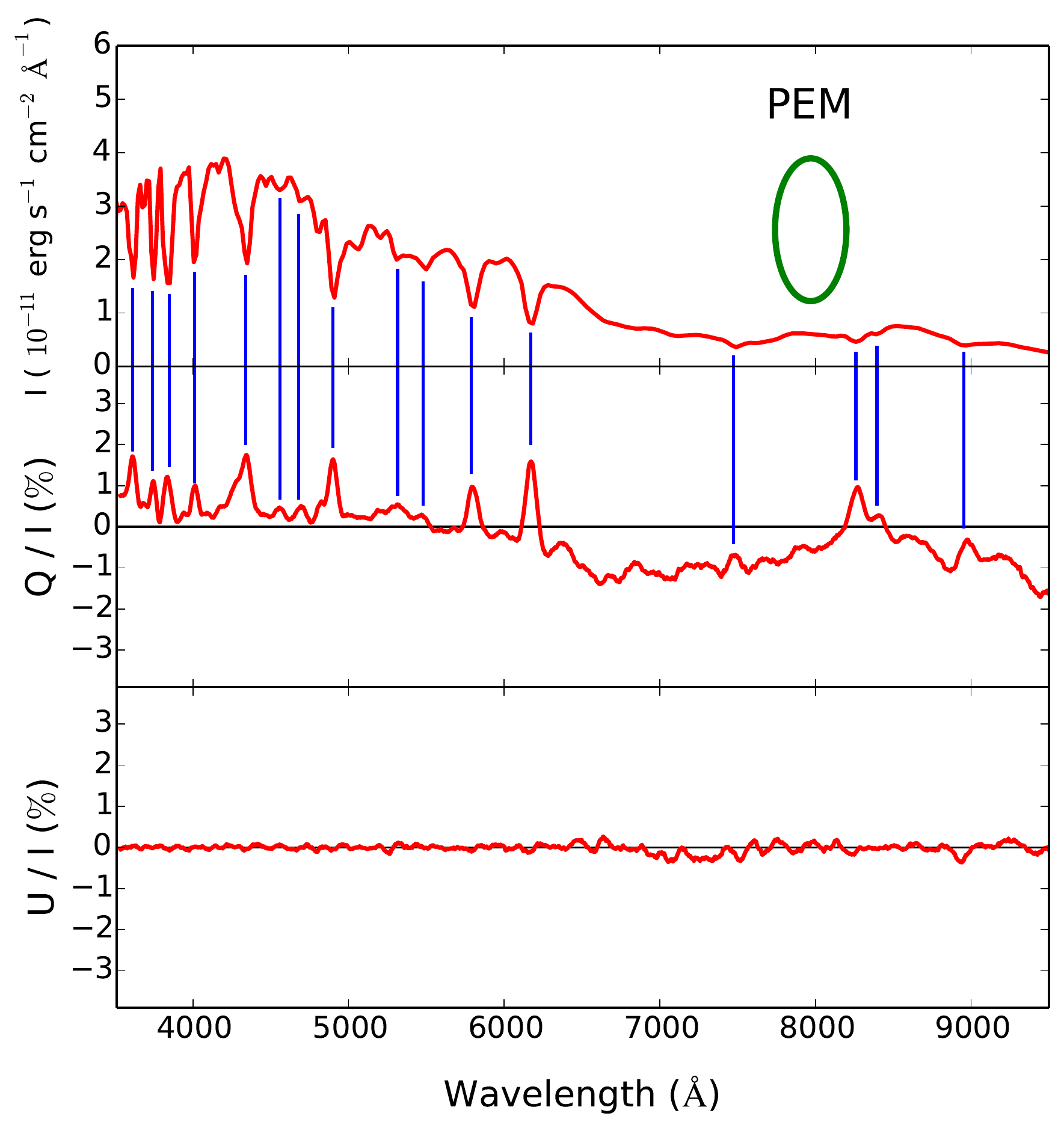}
\hspace{0.5cm}
\includegraphics[width=0.43\textwidth]{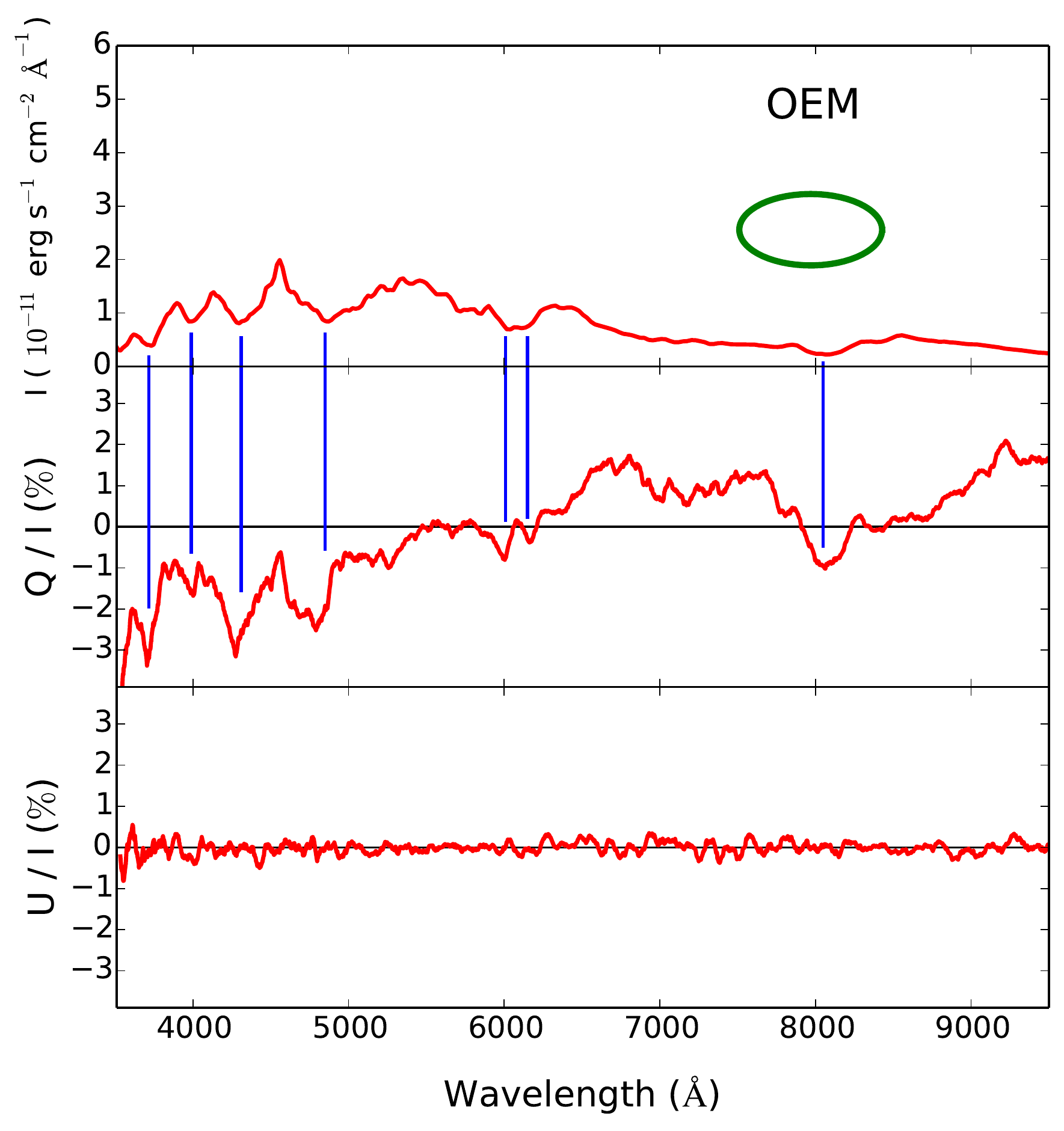}
\caption{Flux and polarisation spectra at $19$ days after explosion for the PEM (left panels) and the OEM (right panels). The observer is placed at $\bmath{n_\text{obs,2}}$ (along $x$). Identification between polarisation features and lines in the spectrum are shown with blue vertical lines. Scaled projected surfaces are shown in green.}
\label{ellipse_pol}
\end{figure*}

\begin{figure*}
\includegraphics[width=0.77\textwidth]{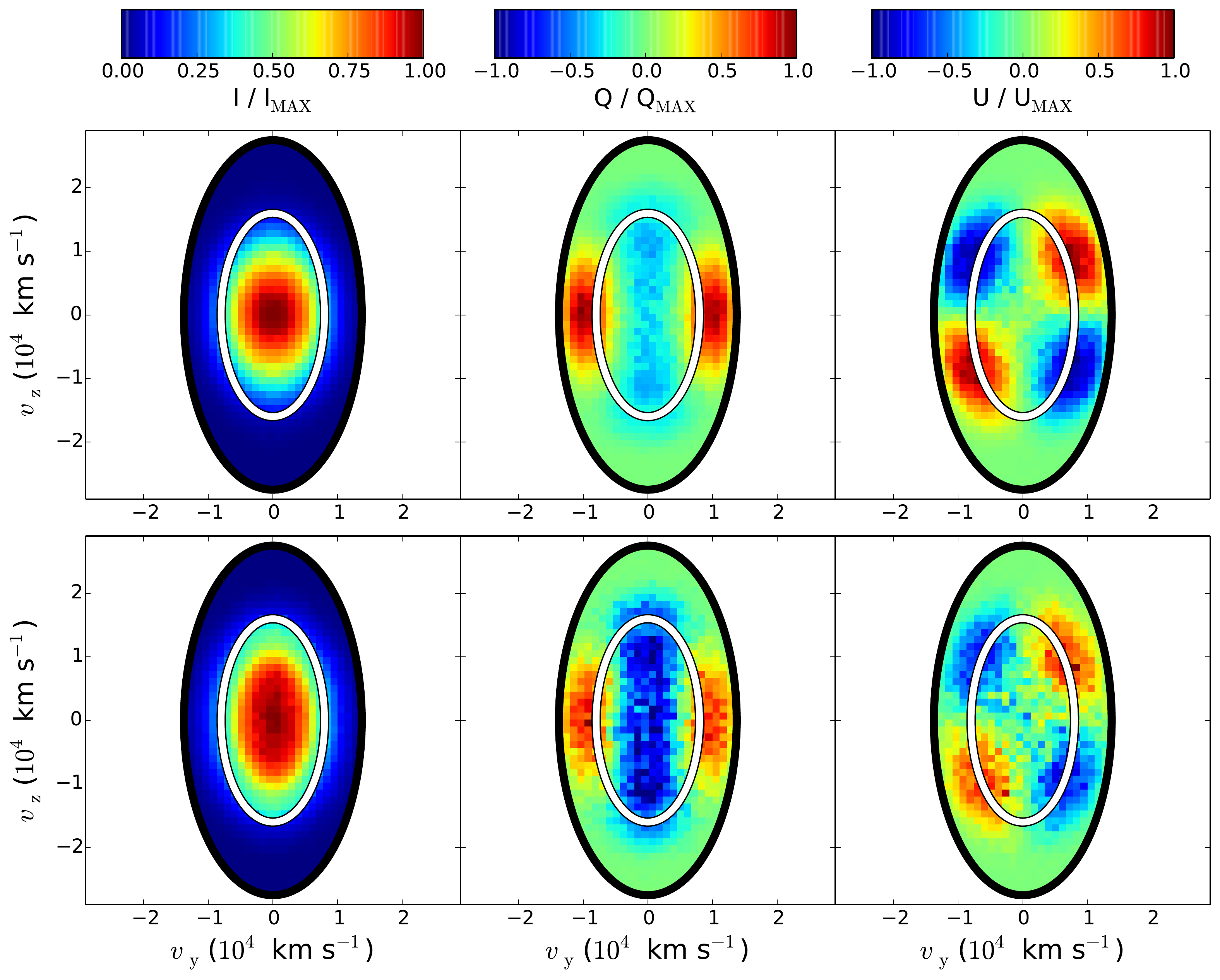}
\caption{Colour maps of normalised $I$ (left panels), $Q$ (middle panels) and $U$ (right panels) distributions projected on the velocity plane ($v_\text{y}$,$v_\text{z}$). The maps are computed for the PEM using the EBT and selecting the emergent $v-$packets between $16.5$ and $21.5$ days after explosion and in the wavelength regions $3500-6000$~\AA{} (upper panels) and $6400-7200$~\AA{} (lower panels). Solid lines mark the outer boundary of the \revised{iron-group-element zone} (inner \revised{white} ellipse) and the maximum velocity parameter \revised{$\xi_\text{max}=13~750$~km~s$^{-1}$} (outer \revised{black} ellipse). The intensity distribution in the blue is more circular than the projected density contour: $Q$ is dominated by contributions along the minor axis, leading to a positive polarisation level. In contrast, the intensity distribution in the red is more similar to the projected density contour: $Q$ is dominated by contributions along the major axis and therefore biased towards a negative value.}
\label{gridave}
\end{figure*}

In this section, we follow previous studies \citep[e.g.][]{hoeflich1991,kromer2009,dessart2011} and use ellipsoidal models as a starting point to explore aspherical geometries. We use a model equivalent to that of \citet{kromer2009}, which has a prolate ejecta morphology (PEM) and also consider a similar model with an oblate ejecta morphology (OEM). Specifically, we assume ellipsoidal isodensity surfaces with density profile  
\begin{equation}
\rho(\xi) \propto \begin{cases} ~\exp\Big(-\frac{\xi}{\xi_0}\Big)  \hspace{0.8cm} \xi < \xi_\text{max} \vspace{0.2cm}\\ \hspace{0.2cm}  0 \hspace{2.25cm} \xi > \xi_\text{max}  \end{cases} ~~~~.
\end{equation}
The parameter $\xi$ is defined in terms of the components of velocity in cylindrical polar coordinates $v=(v_\text{r},v_\text{z})$ as
\begin{equation}
\xi = \begin{cases} ~ \sqrt{~\Big(\frac{v_\text{r}}{h}\Big)^2+v_\text{z}^2}  & \hspace{0.4cm}$OEM$  \vspace{0.4cm}\\ ~ \sqrt{~v_\text{r}^2+\Big(\frac{v_\text{z}}{h}\Big)^2} & \hspace{0.4cm}$PEM$  \end{cases} ~~~,
\end{equation}
where $\xi_0=2750$~km~s$^{-1}$ and $h$ is the ratio between the semi-major and semi-minor axis. Here we limit our study to an axis ratio $h=2$ and fix the maximum \revised{velocity parameter $\xi_\text{max}=13~750$~km~s$^{-1}$}. We adopt a composition for both models that is roughly appropriate for SNe Ia. Specifically, the total mass and chemical yields of the ejecta are chosen to be the same as for the W7 model and a stratified composition with three \revised{ellipsoidal zones ($h=2$) is assumed. The model is set up by filling the ejecta from the centre to accommodate the W7 yields of the different element groups:} the innermost region is filled with ``iron group'' elements ($21\leq Z\leq 30$), the middle with intermediate-mass elements ($9\leq Z\leq 20$) and the outermost with low-mass elements ($Z\leq 8$). \revised{The transitions between the different layers are at $\xi\sim8000$ and $\xi\sim10~500$~km~s$^{-1}$}. Relative abundances of the elements inside each zone are kept fixed to the W7 values. 

LTE radiative transfer calculations have been performed over $111$ logarithmic time-steps from $2$ to $120$ days after explosion. \revised{$2.4\times10^8$} and $4\times10^7$ Monte Carlo quanta have been generated for the PEM and OEM, respectively. Since the redder regions of the polarisation spectra are typically noisier due to the lower flux (i.e. fewer Monte Carlo quanta per frequency bin; see Fig \ref{W7pol}), an additional simulation has been carried out for the PEM (OEM), with $8\times10^7$ ($4\times10^7$) Monte Carlo quanta and with the EBT routine called only for $\lambda>6000$~\AA{}\footnote{This cut in wavelength considerably speeds up the calculation since the $v-$packet routine is called a factor of $\sim25$ times fewer compared to calculations with the entire range $3500-10\,000$~\AA.}. \revised{The two simulations have thus been combined to produce final spectra for the EBT in the whole range between $3500$ and $10\,000$~\AA}. Spectra are computed with the EBT from $10$ to $30$ days after explosion and for two extreme viewing angles: along the $z$-axis, $\bmath{n_\text{obs,1}}=(0,0,1)$, and along the $x$-axis, $\bmath{n_\text{obs,2}}=(1,0,0)$.

\subsubsection{Flux and polarisation spectra}

\begin{figure}
\includegraphics[width=0.482\textwidth,trim=10pt 0pt 5pt 0pt]{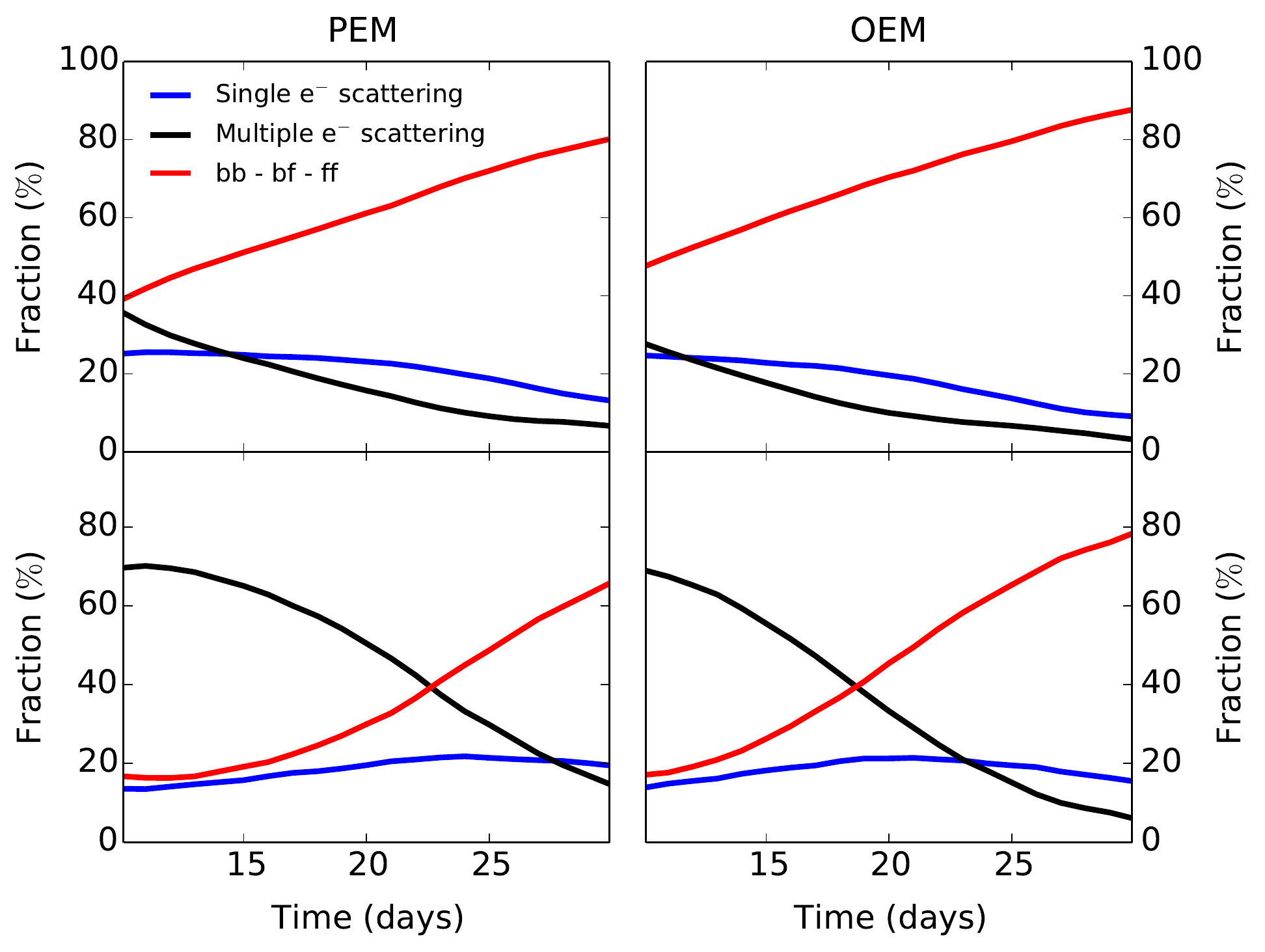}
\centering
\caption{\revised{Evolution of the relative contribution of different scattering processes to the observed spectrum}. Fractions are calculated for the PEM (left panels) and the OEM (right panels) with the DCT by selecting escaping packets based on their last interaction(s) prior to escape. The fraction of packets that underwent a depolarising interaction process (bound-bound, bound-free or free-free emission) as last interaction is shown in red. The contribution from packets that had a single electron scattering interaction since their last depolarising interaction is indicated in blue, and packets that suffered multiple electron scattering events prior to escaping are show in black. Upper panels show contribution in the spectral region $3500-6000$~\AA, lower panels in the wavelength range between $6400$ and $7200$~\AA. }
\label{albedo}
\end{figure}

In Fig. \ref{ellipse_spectra} we compare the $v-$packet total flux spectra at $19$ days after explosion for the two ellipsoidal models. We find the same strong viewing-angle dependencies reported by \citet{kromer2009}. For a given morphology, packets escaping along the major axis see a velocity twice as large compared to the minor axis and the corresponding spectrum is therefore characterised by broader features and stronger line blending; moreover, the spectrum viewed along the major axis is fainter since the projected area along this axis is smaller and the typical opacity is higher. The same geometrical arguments can also be used to compare spectra for the two different geometries: spectra viewed down the minor (major) axis are qualitatively similar, because packets see the same velocity range, but the prolate ellipsoid is fainter than the oblate due to the smaller projected surface.

Polarisation spectra for the observer orientation $\bmath{n_\text{obs,1}}$ are consistent with zero for both models, reflecting the overall spherical symmetry of the projected surface. As shown in Fig. \ref{ellipse_pol}, however, observer orientations from which the model has an elliptical projected surface produce a clear polarisation signal in $Q$. $U$ remains consistent with zero because the model is axi-symmetric, and the calculated $U$ spectrum can be used as a convenient proxy for the Monte Carlo noise in the $Q$ spectrum. 

Sign reversals from shorter to longer wavelengths are found in the $Q$ spectrum for both the PEM and the OEM, a behaviour that can not be explained by the simple picture of an optically thin electron scattering atmosphere illuminated by a point source. In the latter, one would expect the overall polarisation to be negative (positive) for the PEM (OEM), with a polarisation decrease across the lines because of flux dilution. Instead, as found by previous studies \citep{dessart2011,patat2012}, the results of full calculations are more complex and sign reversal in polarisation spectra can arise. These complexities can be ascribed to variations in thermalisation depth with wavelength (see below for explanation) and highlight the need for realistic calculations beyond simple toy atmosphere geometries for the interpretation of data.

\begin{figure}
\includegraphics[width=0.48\textwidth,trim=0pt 0pt 0pt -10pt]{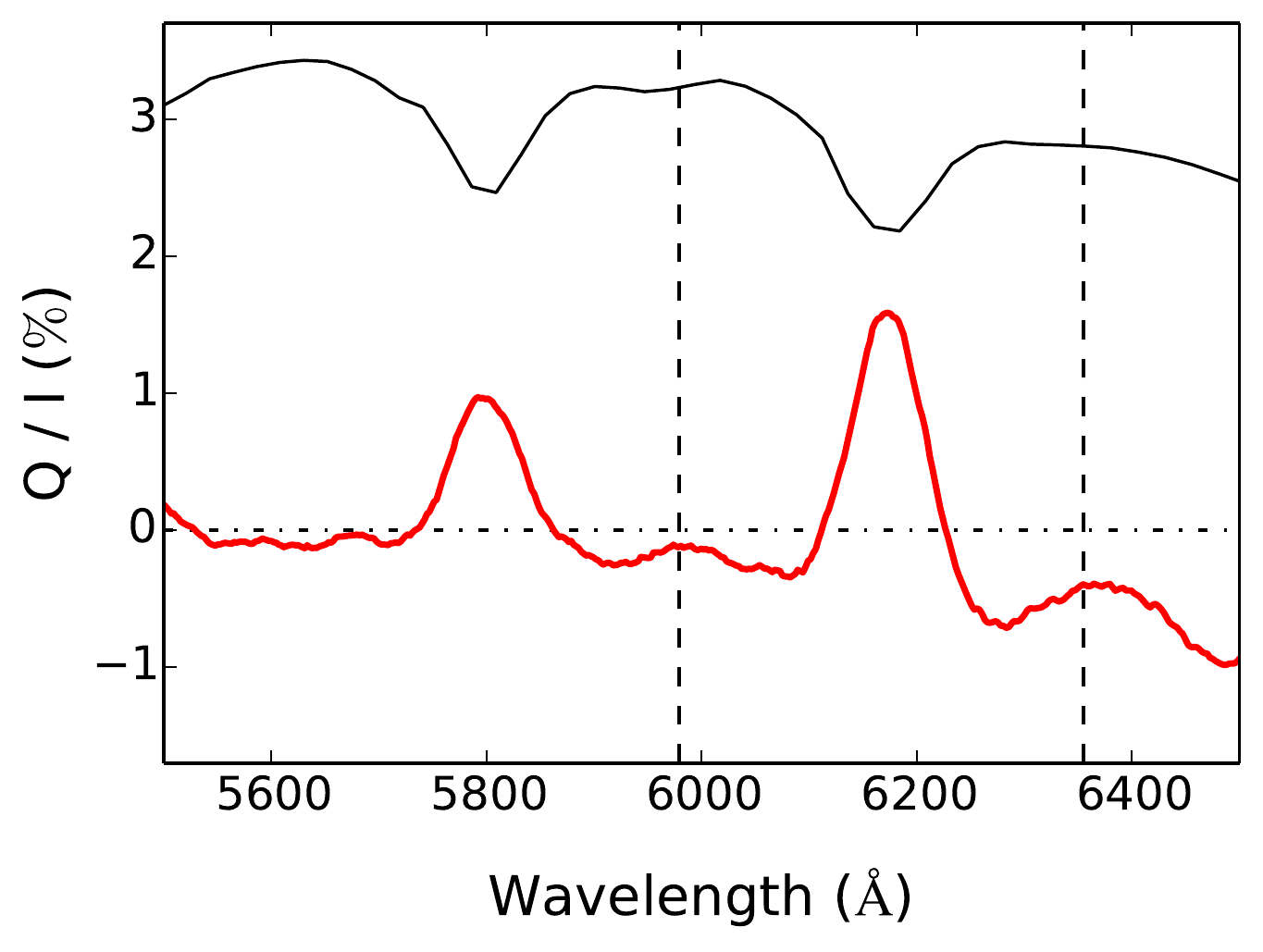}
\centering
\caption{Flux spectrum (solid black line) and $Q$ polarisation spectrum (red line) around the Si\,{\sc ii} $\lambda5979$ and Si\,{\sc ii} $\lambda6355$ features for the PEM viewed along the $x$-axis. \revised{Rest wavelengths of the two lines are marked by vertical dashed lines}. Inverted P-Cygni profiles for the two silicon lines can be identified in the $Q$ spectrum.}
\label{ellipse_silicon} 
\end{figure}

\begin{figure*}
\includegraphics[width=0.49\textwidth]{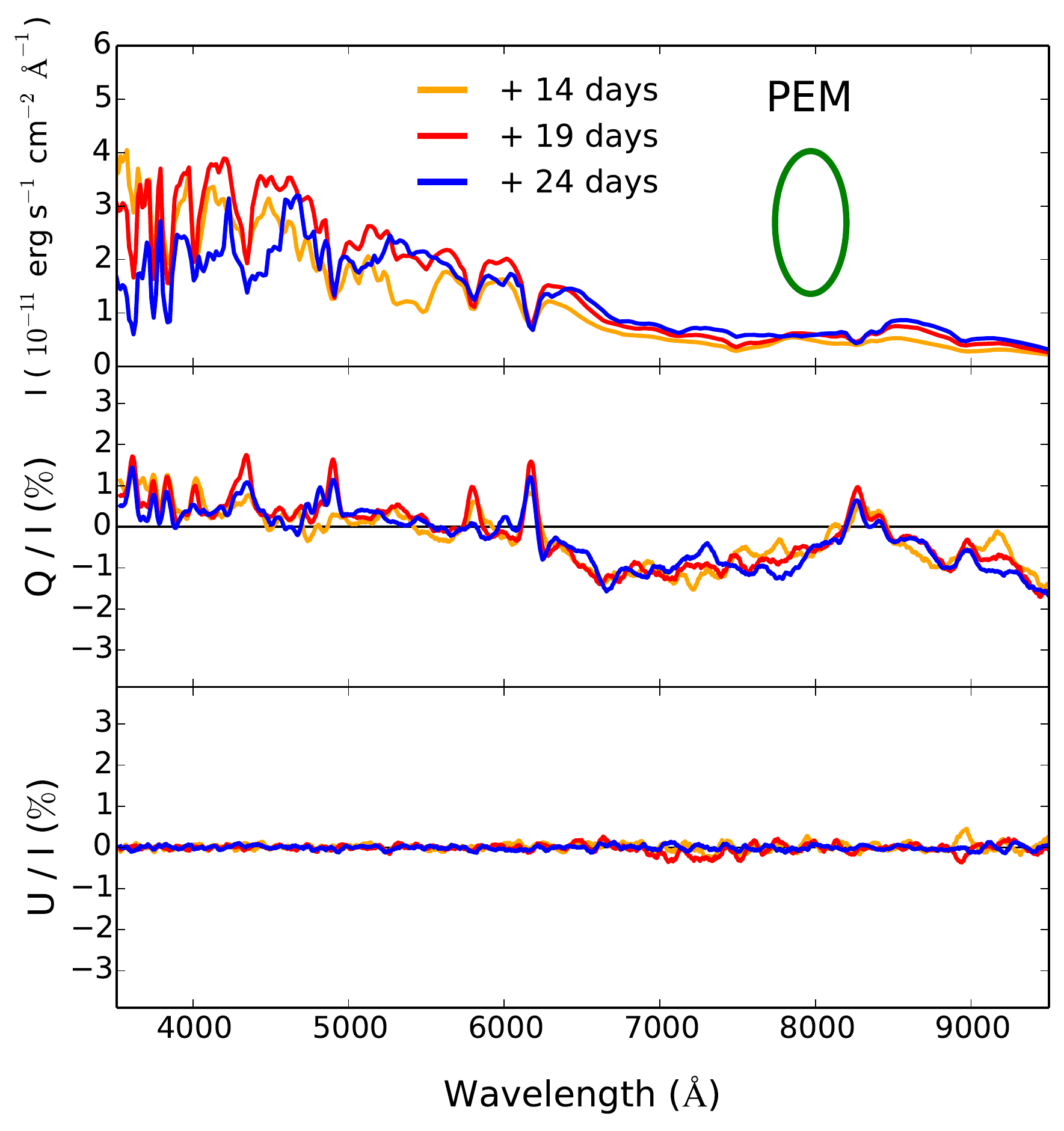}
\includegraphics[width=0.49\textwidth]{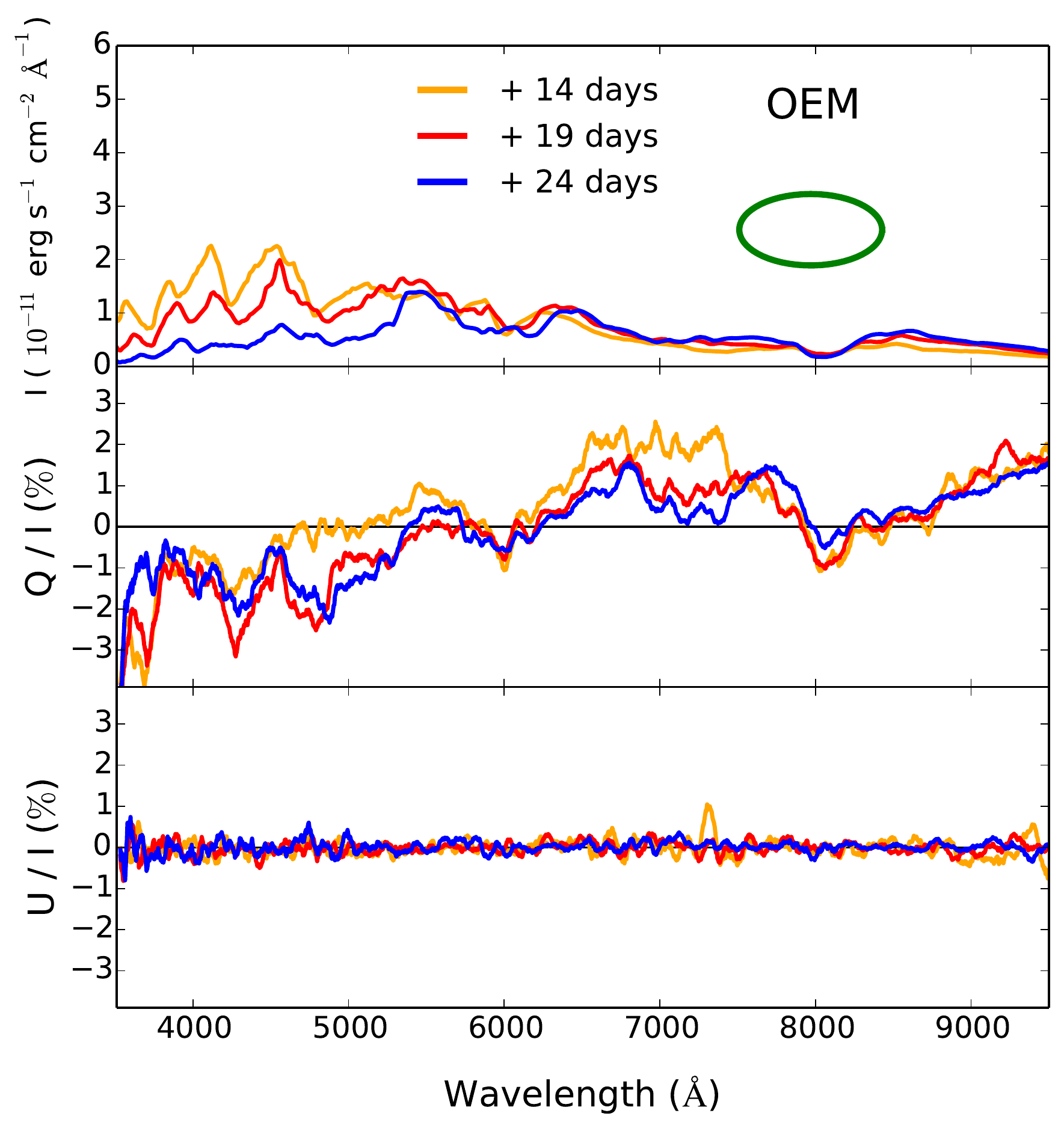}
\caption{Flux and polarisation spectra for the PEM (left panels) and the OEM (right panels) calculated \revised{for a viewing angle $\bmath{n_\text{obs,2}}$ (along $x$) at $14$ (orange), $19$ (red) and $24$ (blue)} days after explosion. Scaled projected surfaces are shown in green. }
\label{ellipse_var}
\end{figure*}

Fig. \ref{gridave} shows the intensity and polarisation distributions projected on the velocity plane ($v_\text{y}$,$v_\text{z}$). The maps have been calculated for the PEM selecting the emergent $v-$packets between $16.5$ and $21.5$ days after explosion and in the spectral regions $3500-6000$~\AA{} and $6400-7200$~\AA. In both wavelength intervals, the intensity emission region in projection is less elliptical than the density contour, and this is a stronger effect in the blue. This behaviour can be ascribed to the relative contributions of the line and the electron scattering opacities in different regions of the spectrum (see Fig. \ref{albedo}): the blue region is dominated by line opacities and thus the intensity distribution in projection is more circular than elliptical\footnote{\revised{This is because, among all the packets created at a given isodensity surface, those at highest projected velocities (i.e. around the major axis of the ellipsoid) sweep out the largest velocity range on their journey to the observer, and therefore encounter the greatest line opacity.}}; in contrast, the red region is free from strong line opacities (see also \citealt{pinto2000}, \citealt{kasen2004} and \citealt{patat2009}) and therefore the projected intensity distribution is more similar to the elliptical density contour. Because contributions to $Q$ are typically positive from regions along the minor axis and negative along the major axis, different distributions in intensity lead to different values of the overall $Q$ polarisation: in the blue, polarisation in $Q$ is dominated by contributions along the minor axis and thus biased towards a positive value, while in the red the $Q$ emission is stronger along the major axis and thus biased toward a negative value. The same arguments explain why the $Q$ polarisation in the OEM changes from negative values in the blue to positive in the red.

The blue region is also characterised by strong ($\sim~1~-~2$~per cent) polarisation across spectral lines: polarisation peaks are associated with the blue-shifted absorption trough, where contributions from the weakly polarised central source are scattered out of the line of sight by the line. In contrast, a decrease in polarisation is found in the emission wing of the P-Cygni profile, where line scattering brings extra unpolarised packets into the line of sight. This leads to the inverted P-Cygni profile shape in the $Q$ spectrum, as discussed by \citet{jeffery1989}. This is clearly visible in the two Si\,{\sc ii} lines at $\sim5979$ and $\sim6355$~\AA{} (see Fig. \ref{ellipse_silicon}).

\subsubsection{Spectral evolution and light curves}

\begin{figure}
\includegraphics[width=0.485\textwidth]{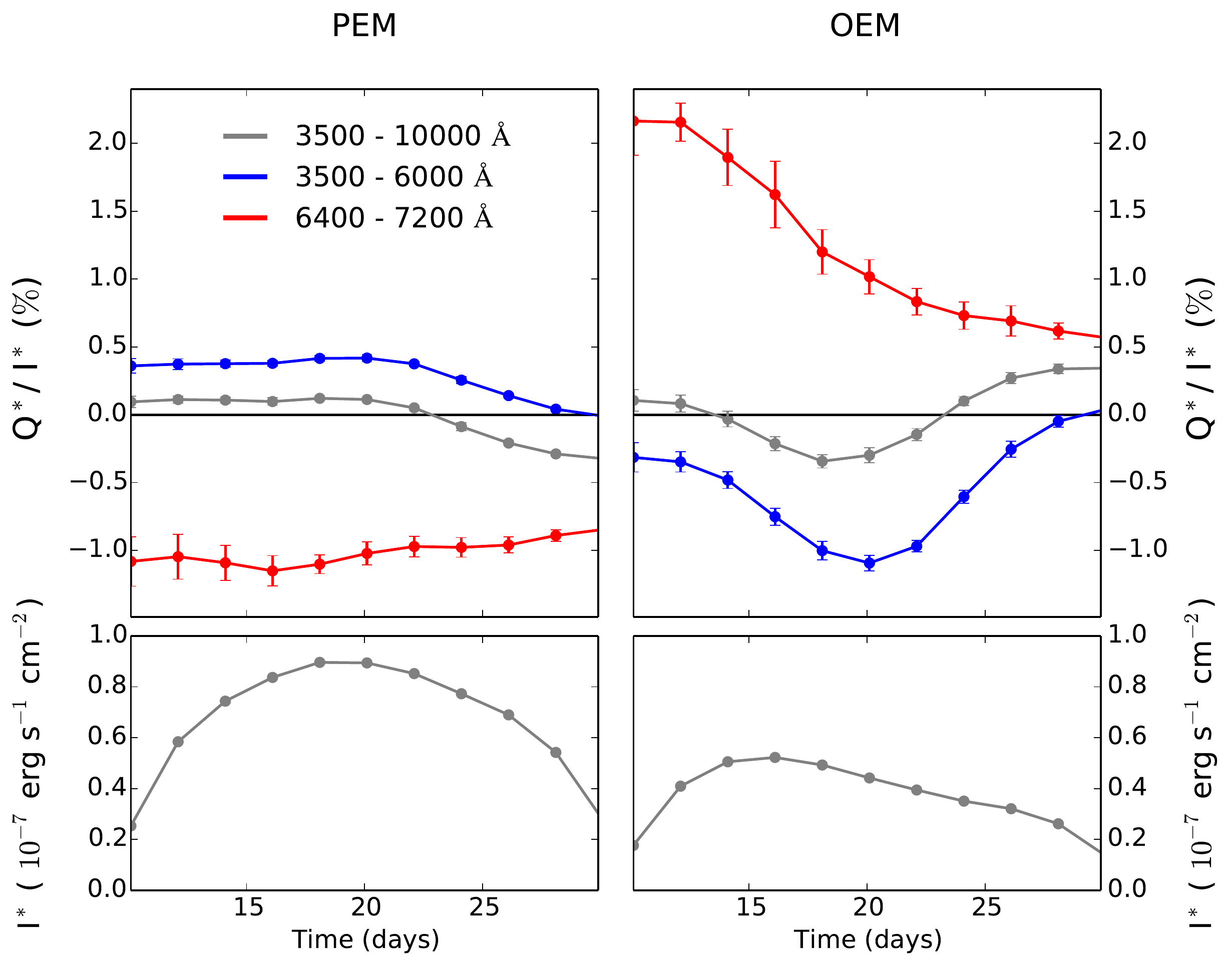}
\centering
\caption{Polarisation light curves between $10$ and $30$ days after explosion for the PEM (left panels) and the OEM (right panels) viewed along the $x$-axis. \revised{Only $Q^*$ is reported here since $U^*$ is consistent with zero for both models}. Grey lines represent contributions from the whole spectral range for which $v-$packets were calculated ($3500-10\,000$~\AA), whereas blue and red lines are for packets escaping at short ($3500-6000$~\AA) or longer\break ($6400-7200$~\AA) wavelengths. Spectral flux integrated in the whole range \revised{is} reported in the \revised{lower} panels. \revised{One-sigma Monte Carlo noise error bars are derived using the procedure outlined in Section \ref{w7}. Some error bars are not visible because they are smaller than the symbol sizes.}}
\label{pol_lc}
\end{figure}

The spectra of both ellipsoidal models are shown in Fig. \ref{ellipse_var} for three epochs ($14$, $19$ and $24$ days after explosion). The integrated luminosity in the wavelength range $3500-10\,000$~\AA{} peaks at $\sim19$ days after explosion in the PEM, whereas the maximum is reached earlier in the OEM ($\sim14$ days after explosion, see Fig. \ref{pol_lc}). To quantify the time evolution of the polarisation, we have also computed polarisation light curves $Q^*(t)$ and $U^*(t)$ by integrating $Q$ and $U$ values over chosen wavelength regions ($\lambda_1$ to $\lambda_2$):
\begin{equation}
Q^*(t) = \int_{\lambda_1}^{\lambda_2}Q(\lambda,t)~\mathrm{d}\lambda \hspace{0.2cm}, \\
U^*(t) = \int_{\lambda_1}^{\lambda_2}U(\lambda,t)~\mathrm{d}\lambda \hspace{0.2cm}.
\end{equation}
As expected, $U^*$ remains consistent with zero at all times. If we consider the entire synthetic spectrum (i.e. $\lambda_1 = 3500$~\AA, $\lambda_2 = 10\,000$~\AA), $Q^*$ in the PEM (OEM) evolves from negative to positive (positive to negative) values as we go from early to late epochs (see Fig. \ref{pol_lc}). This behaviour is due to changes in the relative contributions of the blue and red region as a function of time, and can easily be understood from polarisation light curves computed for the spectral intervals between $\lambda_1=3500$~\AA{} and $\lambda_2=6000$~\AA{} and between $\lambda_1=6400$~\AA{} and $\lambda_2=7200$~\AA{} (see Fig. \ref{pol_lc}). As can be anticipated from the polarisation spectra, $Q^*$ is positive in the blue and negative in the red for the PEM, whereas the opposite is true in the OEM. The OEM light curve evolves more rapidly than the PEM, having reached peak flux at around 14 days and then starting to decline. This more rapid evolution is also clearly evident in the degree of polarisation, which changes much more noticeably over this time period for the OEM. In particular, polarisation in the pseudo-continuum region between $6400$ and $7200$~\AA{} is approximately constant ($Q^* / I^* \sim-1$~per cent) in the PEM, whereas significant evolution is found for the OEM.

\begin{figure}
\includegraphics[width=0.47\textwidth,trim=0pt 0pt 0pt -30pt]{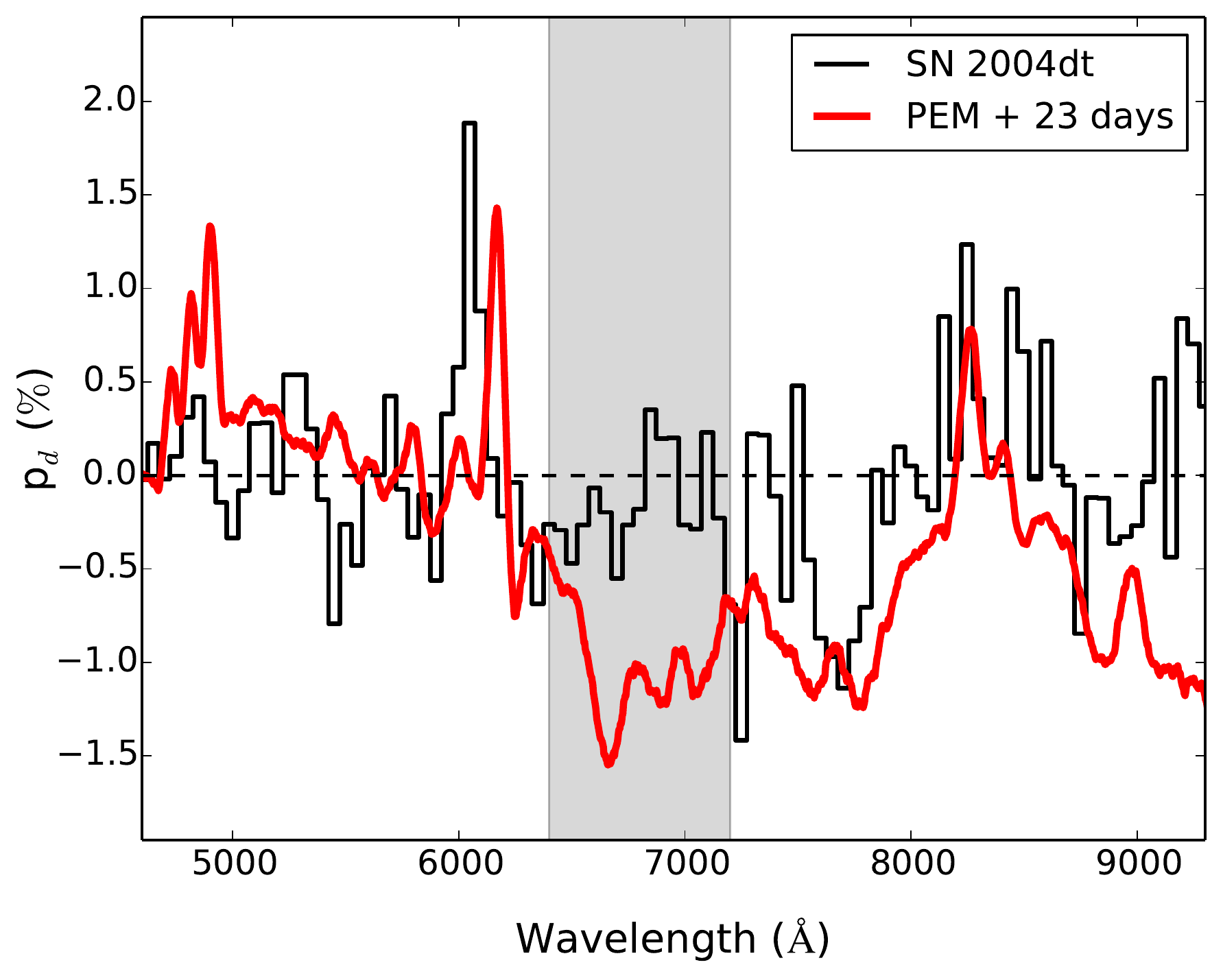}
\centering
\caption{\revised{Q polarisation spectrum for the PEM (in red) calculated $\sim$~4~days after $B$ band maximum light. For comparison the black line shows the polarisation spectrum of SN 2004dt \citep{leonard2005} at the same epoch. Given that the PEM is axisymmetric, the polarisation spectrum of SN 2004dt calculated along the dominant axes, $p_d$, is shown here. Polarisation levels predicted by the PEM across the lines are comparable (within a factor of two) to those observed, while the polarisation in the continuum (grey shaded area) is too high.}}
\label{obscomparison}
\end{figure}

\revised{Although the ellipsoidal models studied here are not very realistic, they do qualitatively reproduce the main features of SN Ia polarisation spectra. As shown in Fig. \ref{obscomparison}, the PEM predicts an overall small polarisation signal throughout the spectrum ($p~\lesssim1.5$~per cent) and polarisation levels across the lines comparable (within a factor of two) to those observed in SN Ia. Of course, the comparison with data is far from perfect: the polarisation predicted in the continuum is too high (because of the strong asymmetry in the toy model) and the velocities of the line features are too small. Such discrepancies come as no surprise, given the simplicity of the model. In future studies, we will make quantitative comparisons to data with results from polarisation calculations performed for real explosion models.}

\section{Conclusions}
\label{conclusions}
We have described a technique for modelling polarisation for multi-dimensional supernova explosion simulations, and implemented it in the radiative transfer code ARTIS \citep{kromer2009}. Our method uses an approach inspired by \citet{lucy2005}, and related to those used by \citet{long2002}, \citet{sim2010} and \citet{kerzendorf2014}, for extracting observables: viewing-angle spectra are obtained by summing contributions from $v-$packets generated at each $r-$packet interaction point and escaping to a chosen observer orientation (Event-Based Technique, EBT). These escaping $v-$packets can be used to construct synthetic observables (spectra, light curves, polarisation spectra) that have substantially reduced Monte Carlo noise compared to spectra obtained by direct binning of escaping $r-$packets. We also investigated a higher-order Monte Carlo noise reduction approach, based not on $r-$packet interaction sites but on $r-$packet trajectory elements (Trajectory-Based Technique, TBT). Initial results, however, suggest that this approach may be less computationally efficient.

We validated our polarisation scheme using an idealised test code in a simple configuration, and found good agreement with predictions from \citet{hillier1994}. Applying the same idealised test code to a simple ellipsoidal toy model, we then verified that continuum polarisation levels calculated with the EBT agree with values predicted by direct binning \revised{of} the escaping $r-$packets. We implemented the EBT in ARTIS and tested it for a model with a realistic SN Ia composition and opacity (the spherically symmetric W7 model): as expected, the $v-$packet method could accurately reproduce the synthetic spectrum obtained by direct binning of emergent Monte Carlo quanta and also predict polarisation consistent with zero. However, the EBT is much less affected by Monte Carlo noise (with typical signal-to-noise ratios a factor of $\sim13$ higher than those obtained with the direct binning approach) and thus more suitable to reproduce very weak signals (e.g. polarisation levels observed in SNe Ia).

Finally, we synthesised flux and polarisation optical spectra with the EBT for prolate and oblate ellipsoids with axis ratio of two, using typical SNe Ia velocities and compositions (including composition layering). As expected, we obtained null polarisation spectra when the projected surface on the plane of the sky is circular. In contrast, aspherical projected areas yield a polarisation signal (typically $\sim1$~per cent) in both morphologies. The polarisation is characterised by sign reversals across the spectrum and peaks associated with troughs of strong optical features. This behaviour is consistent with results of previous studies using similar ejecta morphologies \citep{hoeflich1991,dessart2011,patat2012} and is ascribed to variations in thermalisation depth with wavelength. At the epochs we studied ($14-24$ days post-explosion), the evolution of polarisation spectra is more dramatic for the oblate than the prolate morphology, both in the continuum and in the line polarisation levels. 

\revised{In this paper we have focused on developing our technique and testing its accuracy in calculating intensity and polarisation spectra for one- and two-dimensional models. This study has laid the groundwork for future calculations in which we will exploit the multi-dimensional capability of ARTIS and calculate polarisation spectra for a set of contemporary SN Ia explosion models.} Such calculations will help to identify geometric discriminants between models and to make comparisons between their predictions and spectropolarimetric data more reliable.

\section*{Acknowledgements}

\revised{The authors are thankful to the reviewer, Jennifer Hoffman, for her valuable suggestions which helped to improve the quality of the paper.} 

This work used the DiRAC Data Centric system at Durham University, operated by the Institute for Computational Cosmology on behalf of the STFC DiRAC HPC Facility (www.dirac.ac.uk). This equipment was funded by BIS National E-infrastructure capital grant ST/K00042X/1, STFC capital grant ST/H008519/1, and STFC DiRAC Operations grant ST/K003267/1 and Durham University. DiRAC is part of the National E-Infrastructure.

This research was supported by the Partner Time Allocation (Australian National University), the National Computational Merit Allocation and the Flagship Allocation Schemes of the NCI National Facility at the Australian National University. Parts of this research were conducted by the Australian Research Council Centre of Excellence for All-sky Astrophysics (CAASTRO), through project number CE110001020.

The authors gratefully acknowledge the Gauss Centre for Supercomputing (GCS) for providing computing time through the John von Neumann Institute for Computing (NIC) on the GCS share of the supercomputer JUQUEEN at Jülich Supercomputing Centre (JSC). GCS is the alliance of the three national supercomputing centres HLRS (Universität Stuttgart), JSC (Forschungszentrum Jülich), and LRZ (Bayerische Akademie der Wissenschaften), funded by the German Federal Ministry of Education and Research (BMBF) and the German State Ministries for Research of Baden-Württemberg (MWK), Bayern (StMWFK) and Nordrhein-Westfalen (MIWF).

SAS acknowledges support from STFC grant ST/L000709/1.

\bibliographystyle{mn2e}
\bibliography{bulla2015a}

\end{document}